\documentclass[nofootinbib]{revtex4-2}

\usepackage{graphicx}
\usepackage{animate}
\usepackage{float}
\usepackage[T1,T2A]{fontenc}
\usepackage[utf8]{inputenc}
\usepackage[english]{babel}
\usepackage{amsmath}
\usepackage{lineno}
\usepackage{footmisc}
\usepackage{perpage}
\MakePerPage{footnote}
\usepackage{tikz}
\usepackage{multirow}
\usepackage{caption}
\usepackage{subcaption}
\usepackage{dcolumn}
\usepackage{bm}

\usepackage{pgfplots}
\pgfplotsset{compat=1.15}
\usepackage{mathrsfs}
\usetikzlibrary{shapes.geometric, arrows}

\tikzstyle{startstop} = [rectangle, rounded corners, minimum width=1.8cm, minimum height=1cm,text centered, draw=black, fill=red!30]

\tikzstyle{arrow} = [thin,->,>=stealth]

\addto\captionsenglish{}

\DeclareMathOperator{\arcosh}{arcosh}

\begin{document}

\preprint{APS/123-QED}




\title{Effectively flat potential in the Friedberg-Lee-Sirlin model}

\author{Eduard Kim}
  \altaffiliation{Moscow Institute of Physics and Technology, Institutsky lane 9, Dolgoprudny, Moscow region, 141700}
 \email{kim.e@phystech.edu}
 
 \author{Emin Nugaev}
  \altaffiliation{
   Institute for Nuclear Research of RAS, prospekt 60-letiya Oktyabrya 7a, Moscow, 117312}
 \email{emin@inr.ac.ru}
 
 \affiliation{
}





\begin{abstract}

The Friedberg-Lee-Sirlin (FLS) model is a well-known renormalizable theory of scalar fields that provides for the existence of non-topological solitons. Since this model was proposed, numerous works have been dedicated to studying its classical configurations and its general suitability for various physical problems in cosmology, quantum chromodynamics, etc. In this paper, we study how Q-balls in effective field theory (EFT) reproduce non-topological solitons in full FLS theory. We obtain an analytical description of the simplified model and compare results with numerical calculations and perturbation theory. We also study the condensation of charged bosons on the domain wall. A full numerical solution allows us to check the EFT methods for this problem. The latter analysis is based on the application of EFT methods to significantly inhomogeneous configurations. We give an interpretation of the results in terms of the shifted boson mass and the vacuum rearrangement.

\end{abstract}

\maketitle







\section{\label{intro}Introduction}

Effective field theory is a theoretical instrument widely used in modern physics; see \cite{Burgess:2020tbq}. The success of the EFT method depends on the hierarchy of scales in the original theory, i.e., the parameters of the theory provide a hierarchy of lengths or energy scales. Specifically, decoupling allows one to obtain practical results. Using EFT, one can calculate S-matrix elements in a simplified model or find semi-classical solutions, for example, solitons or instantons. The benefit is that, in principle, calculations might be systematically improved without knowledge of the exact theory, even in a non-perturbative regime. The above-mentioned features made this instrument extremely efficient and convenient when applied to phenomenological theories related to quantum chromodynamics \cite{manohar_wise_2000, Shifman:1987rj}, phase transitions in cosmology \cite{PhysRevLett.77.2887, KARSCH1996217}, etc. 

In this paper, we use the EFT technique for stationary problems in classical field theory. In particular, we apply it to the problem of finding non-topological solitons \cite{LEE1992251, shnir_2018, Nugaev_2020}. The Friedberg-Lee-Sirlin (FLS) model \cite{PhysRevD.13.2739} was chosen on the basis of the following advantages: it is a renormalizable boson theory suitable for semi-classical description. Since this is a theory of two scalar fields, we are interested in integrating out the real field and obtaining a simplified single-field theory with Q-balls. When being applied (with possible modifications) to various phenomenological models in cosmology and particle physics \cite{PhysRevD.87.083528, PhysRevD.105.085013, LEE1992251}, the FLS model showed itself as a useful analytical evaluation tool. Any comments on the quality of the EFT derived from the FLS model will be based on an analysis of the field configurations for the EFT and the original theory.   

Let us now discuss the above-mentioned compact classical field object called the Q-ball in more detail. Localized solutions of the nonlinear equations of motion of classical field theory are solitary waves, which are more often called solitons in the physical literature \cite{rajaraman1982solitons, manton_sutcliffe_2004, shnir_2018}. These objects are essentially non-linear and can be studied only in the non-perturbative regime. The existence of solitons can be ensured by two mechanisms: topological and non-topological. The topological mechanism consists, according to the considered theory, in the presence of a non-trivial vacuum structure and the existence of solutions with a conserved non-zero topological charge \cite{rajaraman1982solitons}.
Additionally, the presence in the theory of an unbroken, continuous internal symmetry provides conserved charge $Q$. The set of special conditions for the potential of a complex scalar field leads to the existence of solitons, called Q-balls. Q-balls have been extensively investigated in numerous previous works \cite{10.1063/1.1664693, COLEMAN1985263}. 

Under the assumption of scale hierarchy among parameters of the theory, one can use methods of EFT. In particular, it is possible to integrate out heavy real field in the FLS model to obtain a single field theory with Q-balls. Surprisingly, the resulted effective potential\footnote{In contrast to the Coleman-Weinberg potential \cite{PhysRevD.7.1888}, we are constructing an EFT only at a classical level by using equations of motion to integrate out real field of the FLS model.} is a smooth piece-wise function partially consisting of a massive $(\phi^{\ast}\phi)^{2}$ model near the field origin and a flat potential elsewhere. The integral characteristics of Q-balls in the theory with effective potential reenact those of the FLS model. The same result might be reproduced in the gradient approximation of the theory.

It is known from previous works that the FLS model significantly differs in $(3+1)$ and $(1+1)$ dimensions \cite{Lensky_2001}. In the original paper \cite{PhysRevD.13.2739}, the model was studied in the three spatial dimensions, and non-topological localized solutions were found to be divided into two different branches. In this case, there are no static topological configurations for the real field could be found due to  Derrick's theorem \cite{10.1063/1.1704233}. In $(1+1)$ dimensions, the situation is different and arguably more diverse. Non-topological solitons are presented in only one branch of classically stable field configurations. For this model, Derrick's theorem does not restrict the existence of topological solitons (or domain walls) within the theory, and they were also studied both numerically and by analytical approximations \cite{PhysRevD.11.2950, Lensky_2001, Heeck:2023idx}.
To take into account the interaction of the domain wall and complex scalar field, it is necessary to reconsider the method of the constructing of the effective potential. Indeed, in the case of domain walls, the presence of non-trivial topological charge significantly distinguishes this vacuum from a usual homogeneous vacuum. In order to construct an appropriate effective theory, we applied both methods of EFT and perturbation theory. Obtained results allowed both analytical and numerical study. We found that bosons are in a bound state within the potential of a kink and are able to form condensate.

We construct an effective potential for non-topological solitons of the FLS model in Sec.\ref{effective}. The Q-balls of the resulted theory in $(1+1)$ and $(3+1)$ dimensions are studied in Sec.\ref{1+1} and Sec.\ref{3+1}, respectively. Sec.\ref{domain walls} is dedicated to the analysis of the Q-balls within the domain wall. We provide a discussion of our results in Sec.\ref{outlook}.

\section{Effective potential}\label{effective}

Let us write the Lagrangian\footnote{Throughout the paper, we will use the following metrics: $(+,-,-,-)$ and $(+,-)$ for the $(3+1)$ and $(1+1)$-dimensional theories, respectively} of the Friedberg-Lee-Sirlin model \cite{PhysRevD.13.2739} with two scalar fields: complex field $\phi$ and real field $\chi$

\begin{equation}\label{eq3.1}
    \mathcal{L} = \partial_{\mu}\phi^{\ast} \partial^{\mu}\phi + \frac{1}{2}\partial_{\mu}\chi \partial^{\mu}\chi - V(|\phi|^2,\chi)
\end{equation}
in which potential is in the form of
\begin{equation}\label{pot}
    V(|\phi|^2,\chi)=h^{2}\vert \phi \vert^{2} \chi^{2} + \frac{m^{2}}{2}(\chi^{2}-v^{2})^{2}
\end{equation}

Corresponding equations of motion are

\begin{equation}\label{eq3.2}
    \begin{cases}
        &\partial_{\mu}\partial^{\mu}\phi + \frac{1}{2}\frac{\partial V(|\phi|^2,\chi)}{\partial |\phi|}=0,\\
        &\partial_{\mu}\partial^{\mu}\chi + \frac{\partial V(|\phi|^2,\chi)}{\partial \chi}=0
    \end{cases}
\end{equation}

Theory (\ref{eq3.1}) implies the existence of non-broken $U(1)$ symmetry with corresponding conserving Noether charge and discrete symmetry $\chi \rightarrow -\chi$. The potential with spontaneous symmetry breaking may provide non-zero topological charge in $(1+1)$ dimensions. Following \cite{MONTONEN1976349}, the ansatz for Eq.(\ref{eq3.2}) solutions might be chosen as

\begin{equation}\label{eq3.3}
\begin{cases}
    &\phi(t,\Vec{x})=e^{-i\omega t}f(\Vec{x}),\\ &\chi(t,\Vec{x})=\chi(\Vec{x})
\end{cases}
\end{equation}

Equations Eq.(\ref{eq3.2}) now could be rewritten as

\begin{equation}\label{eq3.4}
    \begin{split}
        &\nabla^{2}f=h^{2}\chi^2f-\omega^{2}f,\\
        &\nabla^{2}\chi=2h^{2}f^{2}\chi+2m^{2}(\chi^{2}-v^{2})\chi
    \end{split}
\end{equation}
From formulation Eq.(\ref{eq3.4}), we can see that the equations of motion of the theory possess a few dimensional quantities, like $\sqrt{h^2v^2-\omega^2}$ from field $\phi$ asymptotic behavior; $h\phi$ and $mv$. 

In order to derive effective potential, we propose two methods that, in the end, will lead to the same result. A feature underlying their similarity arises by assuming the relation between fields masses $m_{\chi}=mv \gg hv=m_{\phi}$ which is true in any dimensions. What this means is that it is possible to integrate out real field $\chi$ by using the bottom equation of (\ref{eq3.2}). Firstly, one can set up hierarchy among quantities introduced above and gradient terms of Eq.(\ref{eq3.4}); in this case, imposing an overwhelming value of $mv$ corresponding to steep changing of field $\chi$, leads to 

\begin{equation}\label{eq3.5}
    \frac{\nabla^{2}\chi}{2m^2v^2\chi}= \left(\frac{h^{2}\vert \phi \vert^{2}}{m^2v^2}+\left(\frac{\chi^2}{v^2}-1\right)\right)\rightarrow 0
\end{equation}
along with the second solution $\chi=0$ performs integration of field $\chi$. In other words, we want to study two different regions: when field $\chi$ is a constant solution and a region of steep change of $\chi$ to vacuum values. The same results can be derived from the gradient approximation in Eq.(\ref{eq3.4}) when 

\begin{equation}\label{grad.approx}
    \nabla^{2} \chi = \frac{\partial V(|\phi|^2,\chi)}{\partial \chi} = 0
\end{equation}

Equations (\ref{eq3.5}-\ref{grad.approx}) lead to two possible cases. Firstly, if field $\chi$ equals zero, then the value of the potential (\ref{eq3.2}) becomes $V_{1}=\frac{m^{2}v^{4}}{2}$ and from Eq.(\ref{eq3.4}) we get that $\nabla^{2}\chi=0$. Secondly, if field $\chi$ does not equal zero, then 

\begin{equation}\label{eq3.6}
    \chi^{2} = v^2 - \frac{h^{2}}{m^{2}}\vert \phi \vert^{2}     
\end{equation}
with corresponding potential $V_{2}=h^{2}v^2|\phi|^2 - \frac{h^{4}}{2m^{2}}|\phi|^{4}$, where we can denote $m_{\phi}=hv$.
The same result can be obtained from the perspective of the quantum mechanical analog of Eq.($\ref{eq3.4}$) $\frac{\nabla^{2}\chi}{2m^2v^2\chi}=\frac{\chi^{2}}{v^2}-1+\frac{h^2}{m^{2}v^{2}}|\phi|^{2}$ as long as we keep the $\chi$ field "mass" term $m^{2}v^2\chi$ to be greater than its momentum ($\nabla^{2}\chi\leftrightarrow \Hat{p}^2\chi$). In this case, the lower equation of Eqs.(\ref{eq3.4}) can be treated as algebraic with two possible outcomes.
    
Written above, it results in constructing a piece-wise effective potential. Two parts of effective potential are linked ($V_{1}=V_{2}$) at the value of field $|\phi_{s}|$, at some radius $R$. As it was discussed in \cite{shnir_2018}, in order for Q-ball to exist, effective potential must be as $V_{eff}=V_{1}$, if $|\phi|>|\phi_{s}|$, and $V_{eff}=V_{2}$, if $|\phi|<|\phi_{s}|$. Consideration of $\vert \phi \vert^{2}$ term contribution in Eq.(\ref{eq3.6}) might appear crucial to having a correct representation of soliton configurations from the model (\ref{eq3.1}). For the comparison between full effective potential and reduced one we will calculate integral characterisitics of Q-balls in both cases.
\begin{center}
  \includegraphics[width=0.7\textwidth]{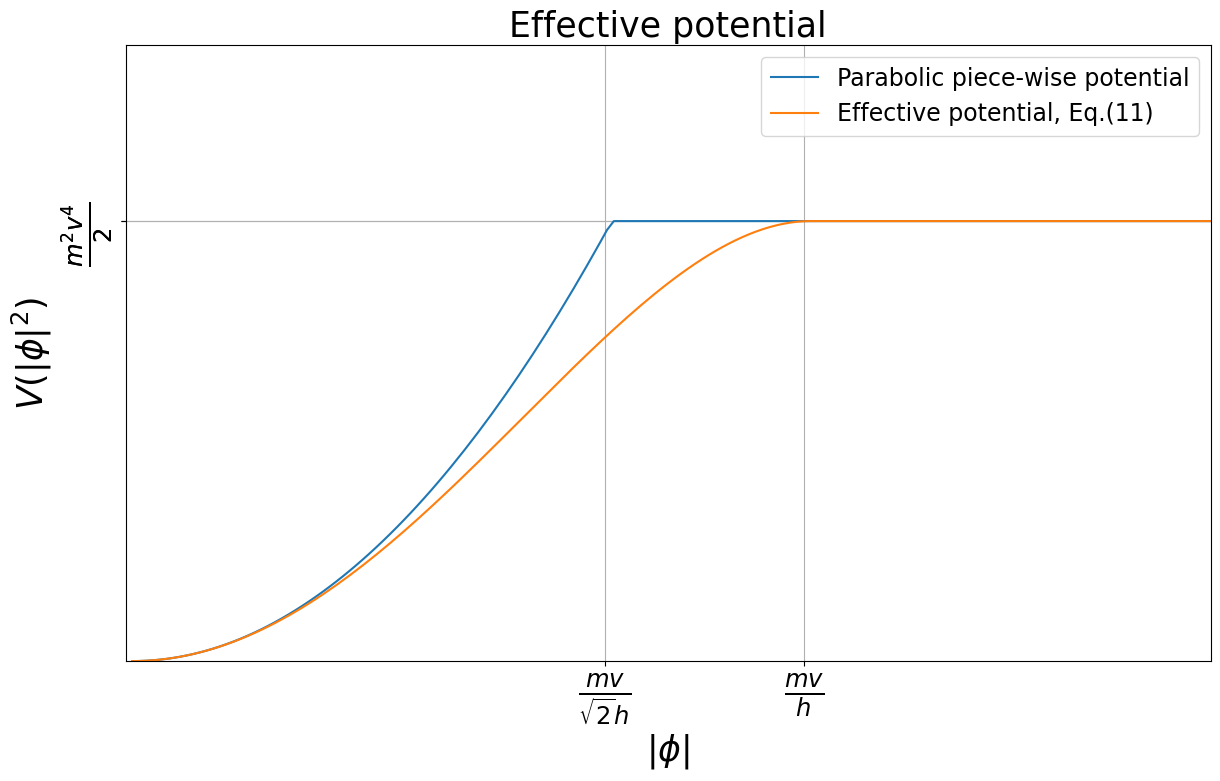}
\captionof{figure}{The effective potential Eq.(\ref{eq3.15}) as the function of $|\phi|$ and reduced parabolic piece-wise potential.}\label{Fig.1}
\end{center}

Thus, one can derive the Lagrangian form of the theory with effective potential

\begin{equation}\label{eq3.14}
    \begin{split}
    &\mathcal{L}= \partial_{\mu}\phi^{\ast} \partial^{\mu}\phi - \left(m_{\phi}^{2}|\phi|^2 -\frac{h^4}{2m^2}|\phi|^{4}\right)\theta \left(\frac{mv}{h}-|\phi|\right)-\frac{m^2v^4}{2}\theta \left(|\phi|-\frac{mv}{h}\right)=\partial_{\mu}\phi^{\ast} \partial^{\mu}\phi-V(|\phi|^2)
    \end{split}
\end{equation}
where $\theta$- is a Heaviside step function. Effective potential is of the form 

\begin{equation}\label{eq3.15}
  V(|\phi|^{2})=\begin{cases}
    m_{\phi}^{2}|\phi|^2-\frac{h^4}{2m^2}|\phi|^4, & \text{if $|\phi|<|\phi_{s}|=\frac{mv}{h}$} \text{(region A)}\\
    \frac{m^2v^4}{2}, & \text{if $|\phi|>|\phi_{s}|=\frac{mv}{h}$ (region B)}
  \end{cases}
\end{equation}

and the equations of motion

\begin{equation}\label{eq3.16}
    \begin{split}
        &A: \nabla^{2}f_{A}(\Vec{x})= (m_{\phi}^{2}-\omega^{2})f_{A}(\Vec{x}) - \frac{h^4}{m^2}f_{A}^{3}(\Vec{x}) \text{, outside of a Q-ball}\\
        &B: \nabla^{2}f_{B}(\Vec{x}) = -\omega^{2}f_{B}(\Vec{x}) \text{, inside the core of a Q-ball}
    \end{split}
\end{equation}

We will use mathematical convention to write $C^{n}$, where n- is the order of the smoothness of the function. In this terms, $V(|\phi|) \in C^{1}$, so at least we would expect $f(\Vec{x})\in C^{2}$. Further investigation of model Eq.(\ref{eq3.14}) in $(1+1)$ and $(3+1)$-dimension space-time will be provided in Sec.\ref{1+1} and Sec.\ref{3+1}.


\section{$(1+1)$-dimensional model}\label{1+1}

In $(1+1)$-dimensional space-time Eqs.(\ref{eq3.16}) are written as

\begin{equation}\label{eq4.1}
    \begin{split}
        &A: f_{A}^{''}(x)= (m_{\phi}^{2}-\omega^{2})f_{A}(x) - \frac{h^4}{m^2}f_{A}^{3}(x)\\
        &B: f_{B}^{''}(x) = -\omega^{2}f_{B}(x)
    \end{split}
\end{equation}

In the reduced single field theory, the number of first integrals of Eqs.(\ref{eq4.1}) allows us to find analytical solution. 
Indeed, since the upper equation in Eqs.(\ref{eq4.1}) is non-linear, the amplitude of the function is strictly fixed by the equation itself. The only possibility remaining is to introduce $x_{0}$ which is the spatial center of the $f_{A}$ solution. The bottom equation of Eqs.(\ref{eq4.1}) is a linear equation with a solution containing multiplier $B_{\omega}$. With typical Q-ball boundary conditions (finiteness of Q-ball's energy requires $\lim_{x \rightarrow \infty}f(x)=0 $ and $\lim_{x \rightarrow \infty}f^{'}(x)=0 $) we find the function $f_{A}$ to be in the form of  

\begin{equation}\label{eq4.2}
    f_{A}(x)=\sqrt{\frac{2m^2(m_{\phi}^{2}-\omega^{2})}{h^4}}\frac{1}{\cosh{\left(\sqrt{m_{\phi}^{2}-\omega^{2}}(|x-x_{0}|)\right)}}
\end{equation}
where $x_{0}$ is the integration constant which corresponds to the center of configuration.

In the region $B$ the solution profile should be even function and it takes form

\begin{equation}\label{eq4.3}
    f_{B}(x)=B_{\omega}\cos{(\omega x)}
\end{equation}

Values of $x_{0}$, $R$ and $B_{\omega}$ are determined from $f_{A}(R)=f_{B}(R)=\frac{mv}{h}$ and $f_{A}^{'}(R)=f_{B}^{'}(R)$ conditions, so that

\begin{equation}\label{eq4.4}
    \begin{split}
        & B_{\omega} = \frac{mv}{h\cos{(\omega R)}}\\
        & x_{0} = R-\frac{\arcosh{ \left(\sqrt{2\left(1-\frac{\omega^2}{m_{\phi}^{2}}\right)}\right)}}{\sqrt{m_{\phi}^{2}-\omega^{2}}}\\
        & R=\frac{\arctan{\left(\sqrt{\frac{m_{\phi}^{2}}{2\omega^{2}}-1} \right)}}{\omega}
    \end{split}
\end{equation}

The determination of the integration constants begets two remarkable things. Firstly, by determining three parameters $B_{\omega}, x_{0} \text{ and } R$ we can directly check that $f_{A}^{''}(R)=f_{B}^{''}(R)$, so $f(x)\in C^{2}$. 
From the solution (\ref{eq4.4}) for the matching radius $R$ one can find restrictions for $\omega$ to be in the interval of $\omega \in \left(0,\frac{m_{\phi}}{\sqrt{2}}\right)$. When $\omega > \frac{m_{\phi}}{\sqrt{2}}$ potential (\ref{eq3.15}) turns into the plain massive $(\phi^{\ast}\phi)^{2}$ potential.

Now we are ready to present values of $U(1)$ charge and energy for Q-ball in the theory (\ref{eq4.1})

\begin{equation}\label{eq4.5}
\begin{split}
    &Q = 2|\phi_{s}|^{2}\left(\frac{\omega R}{\cos^{2}{(\omega R)}} + \tan{(\omega R)} \right) + \frac{8m^2 \omega}{h^4} \left( \sqrt{m_{\phi}^{2}-\omega^{2}}-\sqrt{\frac{m_{\phi}^{2}}{2}-\omega^{2}} \right)
\end{split}
\end{equation}

\begin{equation}\label{eq4.6}
\begin{split}
    &E = \omega Q + 2|\phi_{s}|^{2}\left(\frac{\omega^{2}}{\cos^{2}{(\omega R)}}(R-\frac{\sin{(2\omega R)}}{2\omega}) \right) \frac{4m^2}{3h^4}(m_{\phi}^{2}-\omega^{2})^{\frac{3}{2}}\left(1-\tanh^{3}{\left(\sqrt{m_{\phi}^{2}-\omega^{2}}(R-x_{0})\right)} \right)
\end{split}
\end{equation}
when $\omega \in \left(0, \frac{m_{\phi}}{\sqrt{2}}\right)$. For the rest interval of parameter $\omega$ the results are provided by plain massive $(\phi^{\ast}\phi)^{4}$ theory  

\begin{equation}\label{eq4.7}
    Q = \frac{8m^2\omega}{h^4}\sqrt{m_{\phi}^{2}-\omega^{2}}
\end{equation}

\begin{equation}\label{eq4.8}
    E = \frac{8m^2}{h^4}\sqrt{m_{\phi}^{2}-\omega^{2}}\left(m_{\phi}^{2}-\frac{2}{3}(m_{\phi}^{2}-\omega^{2}) \right)
\end{equation}

One can check that the following differential relation

\begin{equation}\label{Legendre}
    \frac{dE}{d\omega}=\omega\frac{dQ}{d\omega} \text{ or } \frac{dE}{dQ}=\omega
\end{equation}
is satisfied. Eq.(\ref{Legendre}) is fulfilled in a majority of theories with $U(1)$ symmetry.

\subsection{Parabolic piece-wise potential}\label{parabolic1+1}

Before going further, let us investigate how the presence of the non-linear $(\phi^{\ast}\phi)^2$ term in the effective potential affects its ability to reproduce integral characteristics of the Friedberg-Lee-Sirlin model. Lagrangian in this case is very similar to the one studied in \cite{PhysRevD.87.085043}

\begin{equation}\label{eq3.7}
\begin{split}
    &\mathcal{L}= \partial_{\mu}\phi^{\ast} \partial^{\mu}\phi - h^2v^2|\phi|^2 \theta \left(\frac{mv}{h\sqrt{2}}-|\phi|\right)-\frac{m^2v^4}{2}\theta \left(|\phi|-\frac{mv}{h\sqrt{2}}\right)=\partial_{\mu}\phi^{\ast} \partial^{\mu}\phi-V(|\phi|^2)
\end{split}
\end{equation}
where $|\phi_{s}|=\frac{mv}{h\sqrt{2}}$. On first glance, this linear model seems to be more convenient than Eq.(\ref{eq3.14}), and it provides an analytical solution for models in any dimensions. Potential is a piece-wise function of $|\phi|$ with two main properties.

When $|\phi|<|\phi_{s}|$ (region A) model (\ref{eq3.7}) is a free massive scalar field theory, and if $|\phi| > |\phi_{s}|$ (region B) $V_{eff}$ is a flat potential, which ensures the existence of Q-ball.
At the point $|\phi|=|\phi_{s}|$ $V_{eff}$ remains continuous, and $f$ should be at least $C^{1}$ function.

The equations of motion for $(1+1)$-dimensional space-time are the same as Eqs.(\ref{eq4.1}) with the only difference being in the abundance of non-linear terms. In different regions, the solution takes form

\begin{equation}\label{eq3.9}
    \begin{split}
        & A: f_{A}(x) = \frac{mv}{\sqrt{2}h}e^{\sqrt{m_{\phi}^{2}-\omega^{2}}(R-x)}\\
        & B: f_{B}(x) = \frac{mv}{\sqrt{2}h}\frac{\cos{(\omega x)}}{\cos{(\omega R)}}
    \end{split}
\end{equation}
where $f_{A}(R)=f_{B}(R)=\frac{mv}{\sqrt{2}h}$. The matching radius $R=\frac{\arctan{\left(\frac{m_{\phi}^{2}-\omega^{2}}{\omega} \right)}}{\omega}$ determined from condition $f_{A}^{'}(R)=f_{B}^{'}(R)$. Now $U(1)$ charge and energy of Q-ball could be calculated as functions of the free parameter $\omega$

\begin{equation}\label{eq3.10}
    Q = \frac{m^2v^2}{h^2}\left(\frac{\omega}{\sqrt{m_{\phi}^{2}-\omega^{2}}}+\frac{\omega R}{\cos^2{(\omega R)}}+\tan{(\omega R)} \right)
\end{equation}

\begin{equation}\label{eq3.11}
    E = \frac{m^2v^2}{h^2}\left(\frac{m_{\phi}^2}{\sqrt{m_{\phi}^{2}-\omega^{2}}}+\frac{\omega^{2}R}{\cos^{2}{(\omega R)}} + m_{\phi}^{2}R  \right)
\end{equation}
equations (\ref{eq3.10},\ref{eq3.11}) might be additionally checked with the relation (\ref{Legendre}).

\begin{center}
  \includegraphics[width=0.7\textwidth]{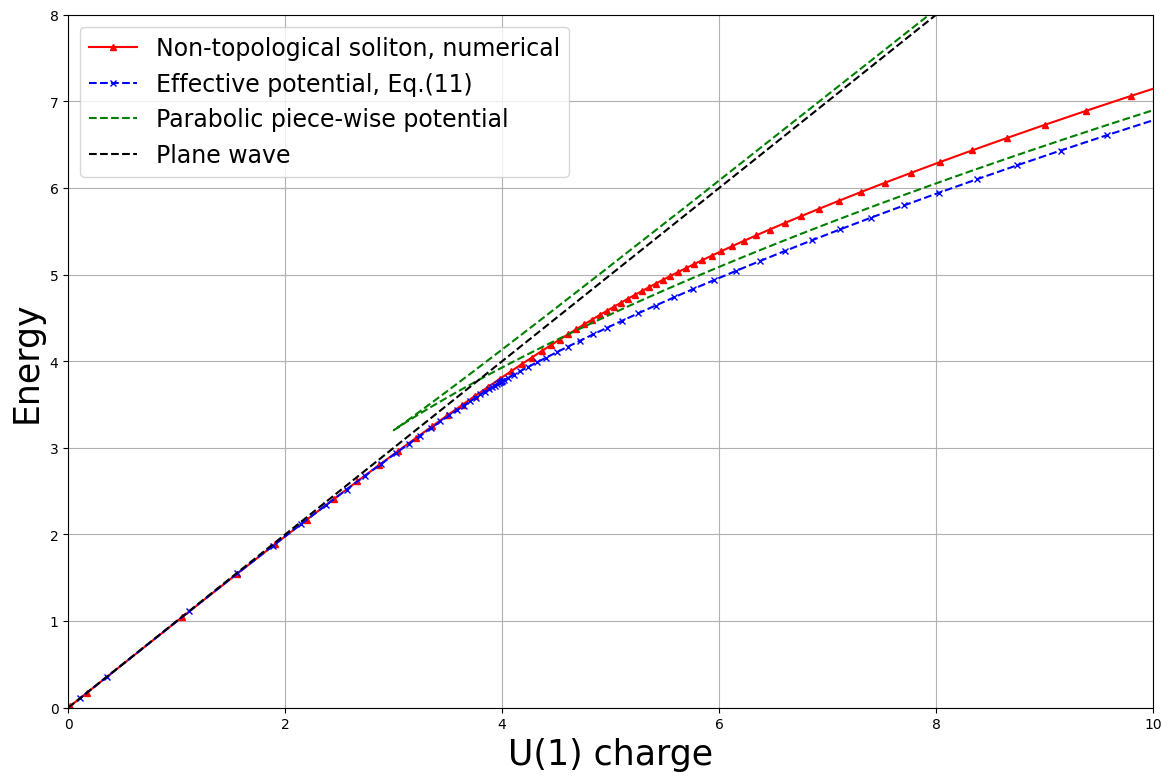}
\captionof{figure}{The energies of the FLS non-topological solitons, effective and parabolic piece-wise potentials Q-balls vs. their $U(1)$ charge plotted for the $(1+1)$-dimensional theory.}\label{Fig.3}
\end{center}

\subsection{Comparison with numerical results}

From the Fig.\ref{Fig.3} one can see that the effective potential and its parabolic approximation reproduce the correct asymptotic for integral characteristics at large charges of stable Q-balls. In order to better understand the mathematical aspects underlying this result, one can see how field profiles from effective potential differ from numerical solutions of Eq.(\ref{eq4.1}). Knowing the analytical solution of the model with effective potential, one can see how field $\chi$ was integrated out

\begin{equation}\label{eq4.9}
\begin{split}
    & A: \chi_{eff}=\sqrt{v^{2}-\left(\sqrt{\frac{2m^2(m_{\phi}^{2}-\omega^{2})}{h^4}}\frac{1}{\cosh{\left(\sqrt{m_{\phi}^{2}-\omega^{2}}(x-x_{0})\right)}} \right)^{2}} \\
    & B: \chi_{eff}=0
\end{split}
\end{equation}

\begin{figure}[H]
\centering
\subfloat{\includegraphics[width=0.4\textwidth]{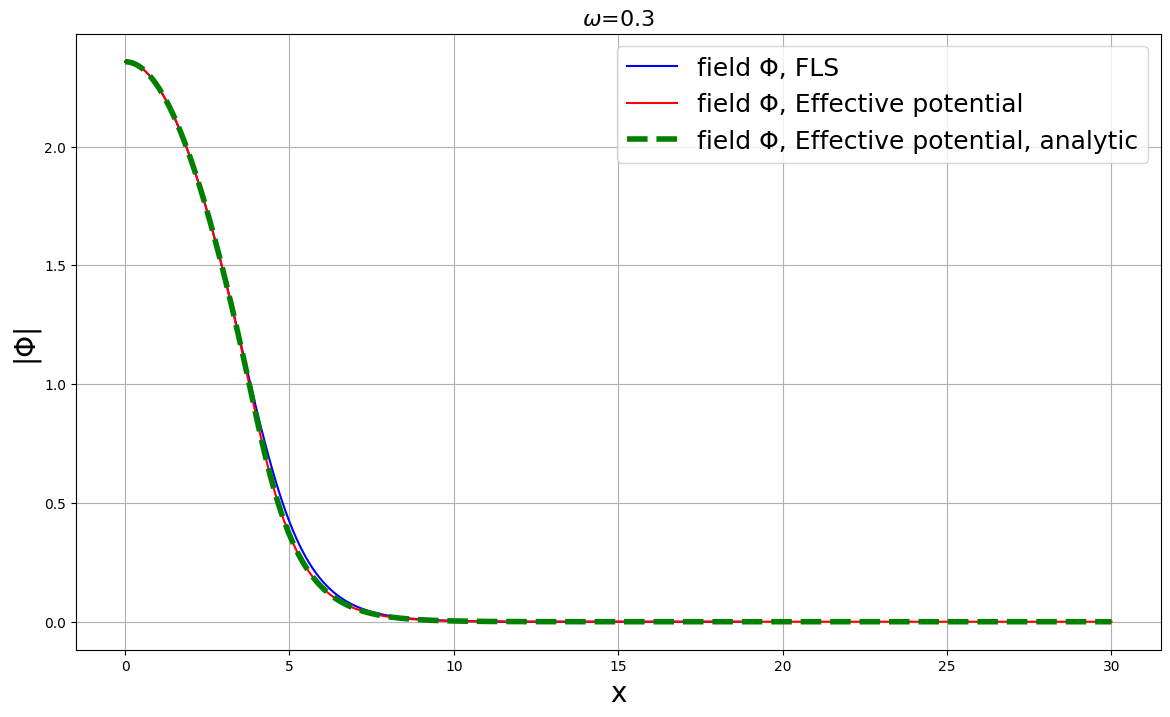}}\hspace{0.5cm}
\subfloat{\includegraphics[width=0.4\textwidth]{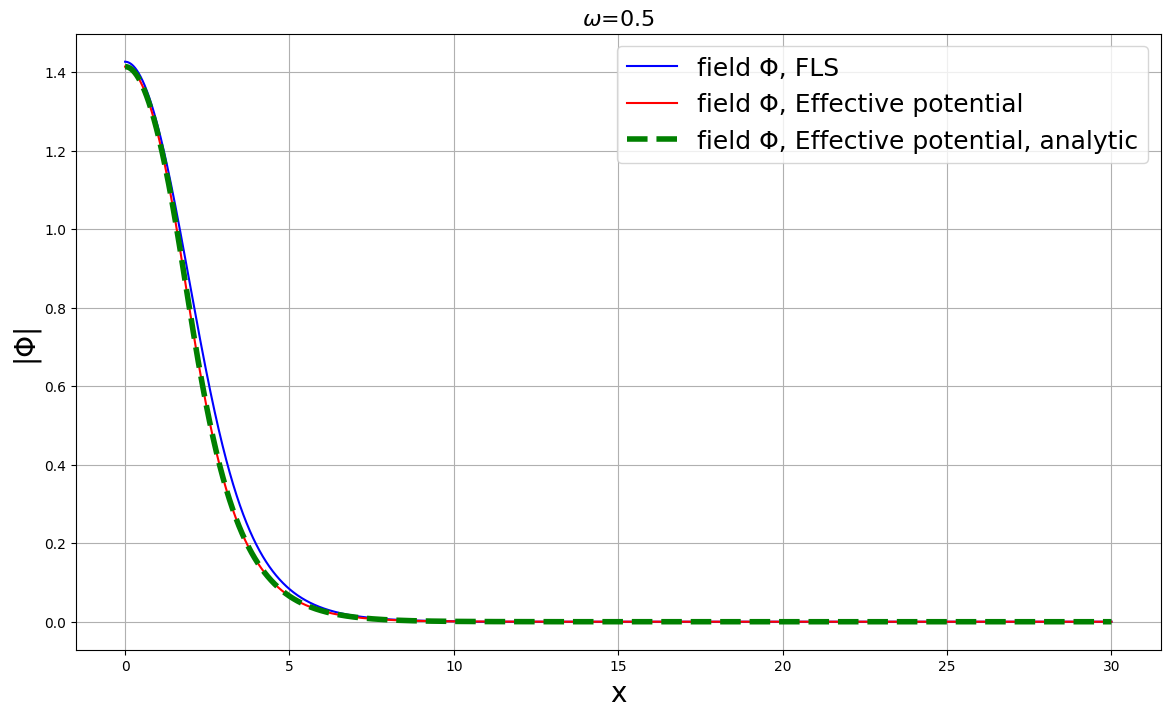}}\\
\subfloat{\includegraphics[width=0.4\textwidth]{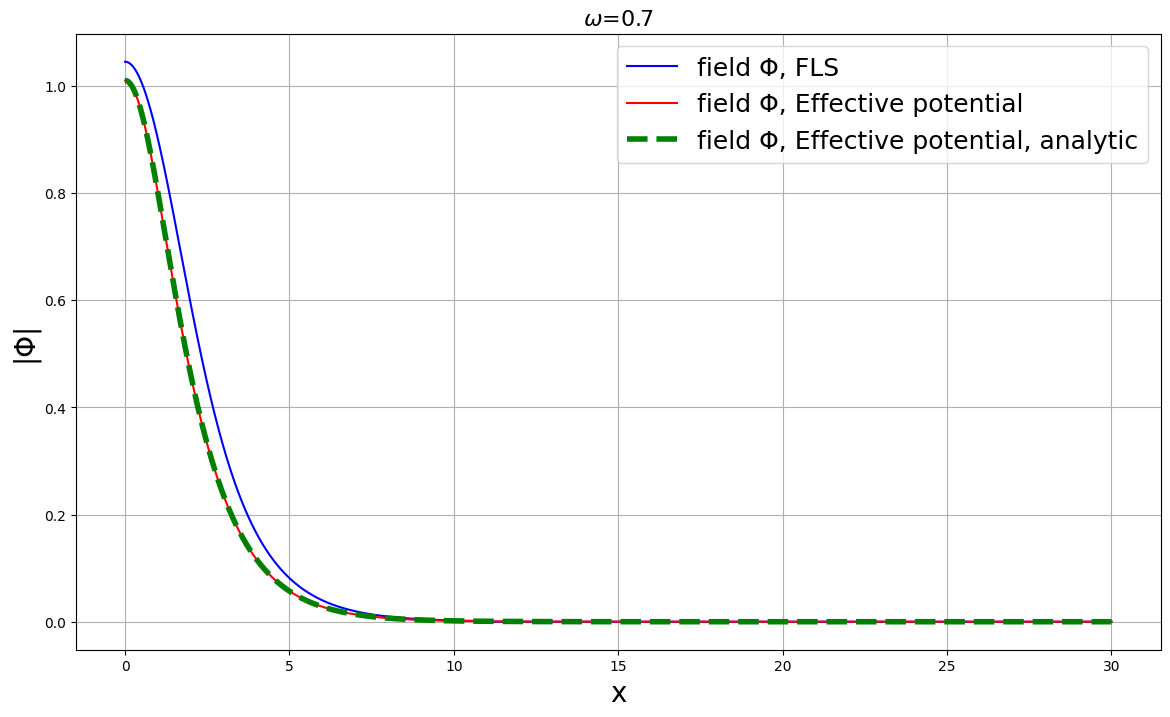}}\hspace{0.5cm}
\subfloat{\includegraphics[width=0.4\textwidth]{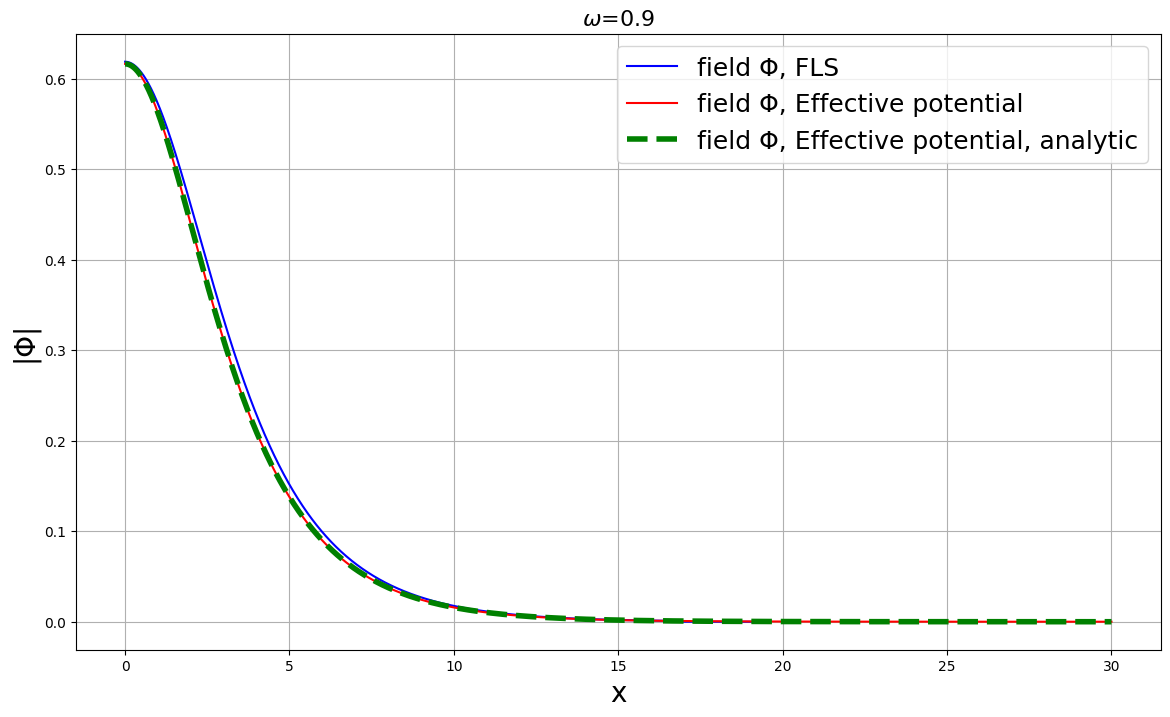}}\\
\caption{The profiles of the field $\phi$ from the FLS model and theory with the effective potential are plotted for different values of parameter $\omega$.} 
\label{f.comparison1}
\end{figure}

\begin{figure}[H]
\centering
\subfloat{\includegraphics[width=0.4\textwidth]{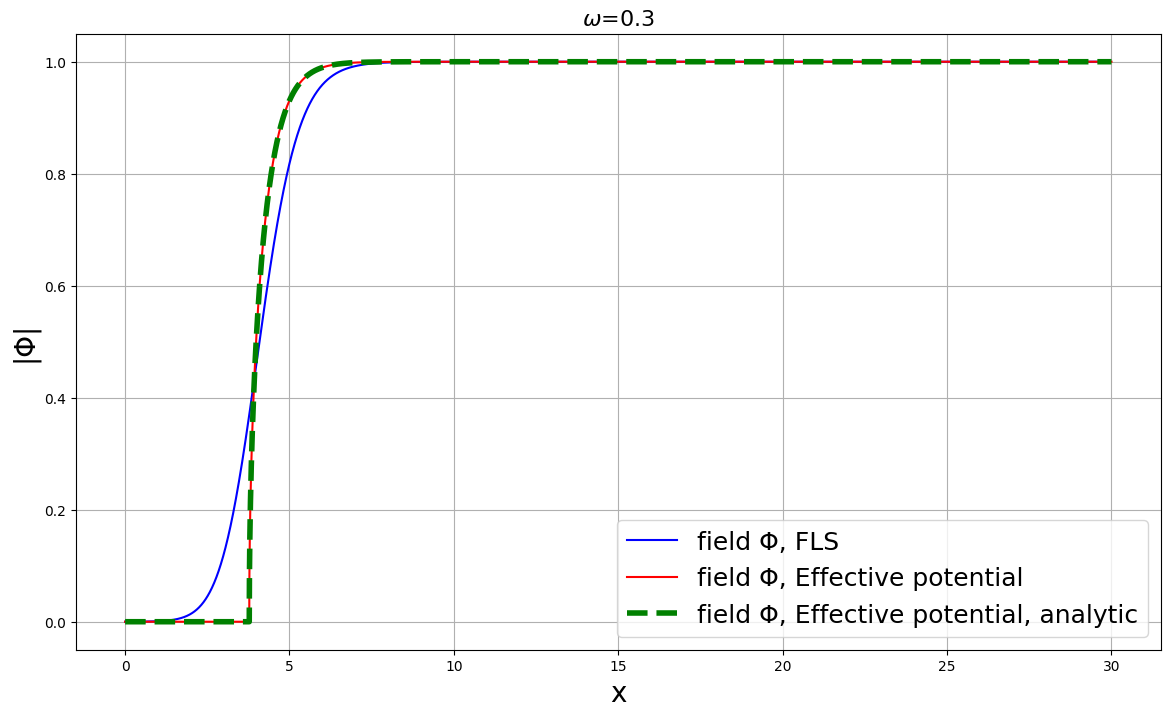}}\hspace{0.5cm}
\subfloat{\includegraphics[width=0.4\textwidth]{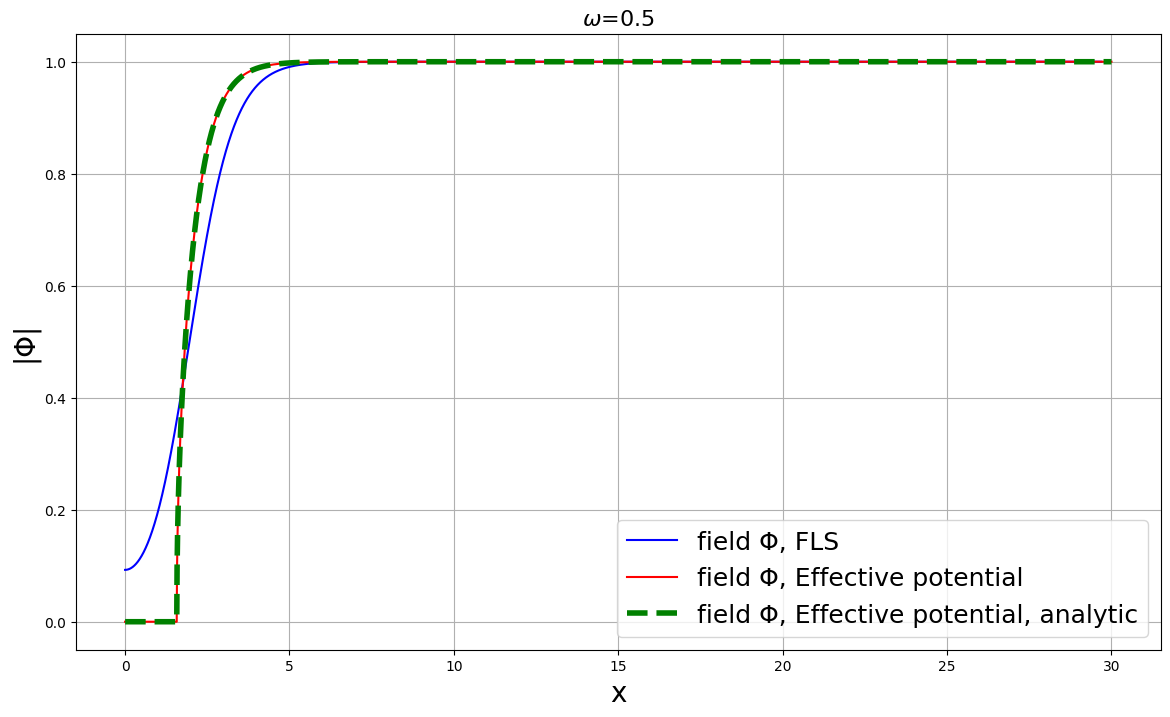}}\\
\subfloat{\includegraphics[width=0.4\textwidth]{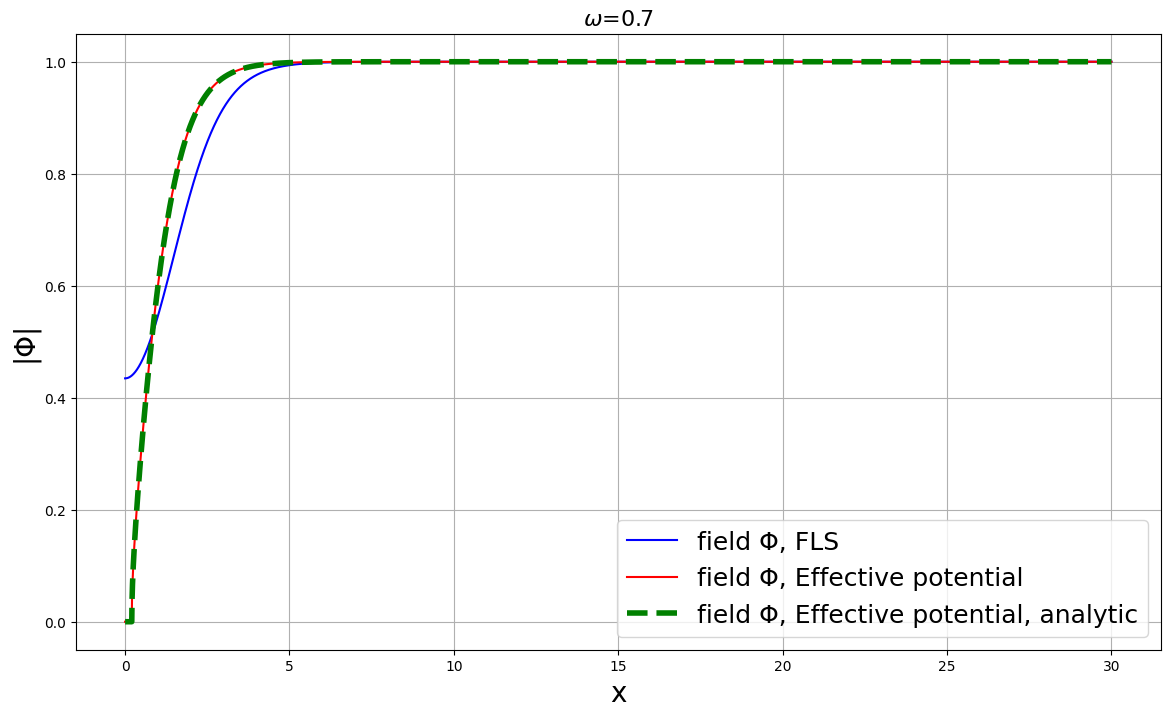}}\hspace{0.5cm}
\subfloat{\includegraphics[width=0.4\textwidth]{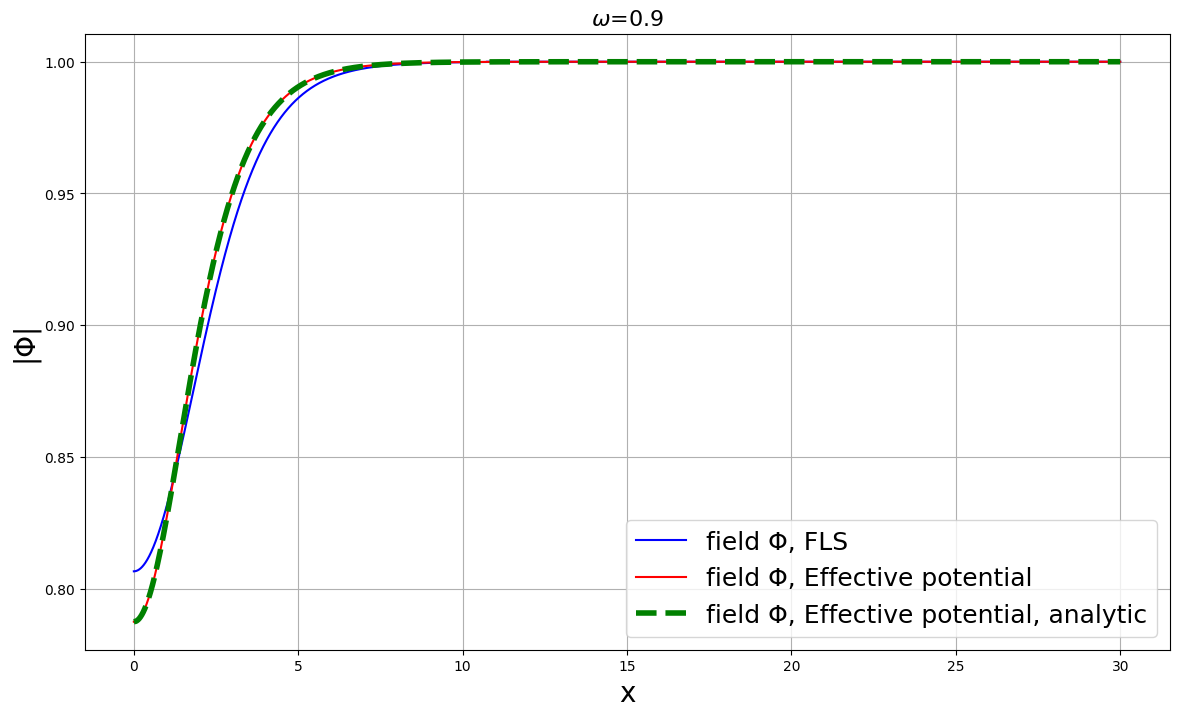}}\\
\caption{The comparison of the field $\chi$ profile from the FLS model and one from the integration procedure Eq.(\ref{eq4.9}) is shown for different values of parameter $\omega$. } 
\label{f.comparison2}
\end{figure}

A feature of Eq.(\ref{eq4.9}) can be seen in Fig.\ref{f.comparison2}, which is the presence of the bubble of the effectively massless field $\phi$. If one takes a close look at the equations above and Fig.(\ref{f.comparison1},\ref{f.comparison2}) the inferiority of the full description of the FLS model Eq.(\ref{eq3.1}) through effective potential is seen. Equations (\ref{eq4.9}) does not allow the resurgence of kink-like solutions with non-trivial vacuum structure. Moreover, effective theory reproduces the profile of complex scalar field $\phi$ as shown in the Fig.\ref{f.comparison1}.

The Friedberg-Lee-Sirlin model (\ref{eq3.1}) is known for not only having a soliton stabilized by $U(1)$ charge when both fields $|\phi|$ and $\chi$ are even, but also for having a configurations with non-trivial topology \cite{Lensky_2001}. This model is of interest due to the possibility to apply EFT methods for inhomogeneous configurations, see also Sec.\ref{domain walls}. 

\section{$(3+1)$-dimensional model}\label{3+1}

In $(3+1)$ dimensions equations of motion for a model with effective potential Eq.(\ref{eq3.15}) is in the form of 

\begin{equation}\label{effective.eom(3+1)}
    \begin{cases}
        &A: \partial_{r}\left(r^2 \partial_{r}f_{A}(r) \right)= r^{2}\left[(m_{\phi}^{2}-\omega^{2})f_{A}(r) - \frac{h^4}{m^2}f_{A}^{3}(r)\right]\\
        &B: \partial_{r}\left(r^2 \partial_{r}f_{B}(r) \right) = -r^{2}\omega^{2}f_{B}(r)
    \end{cases}
\end{equation}
where $\partial_{r}\equiv \frac{\partial}{\partial r}$.

As it was mentioned before, field $\phi$ in the region $A$ corresponds to the $(\phi^{\ast}\phi)^2$ theory that was extensively studied before. From \cite{1970JMP....11.1336A}, we know that all particle-like solutions in $(\phi^{\ast}\phi)^2$ theory are classically unstable. Indeed, there is only one branch of solutions that is unstable both classically and kinematically. The difference in effective potential (\ref{eq3.15}) is that for large values of field $\phi$ potential is a flat function. As shown in Fig.(\ref{Fig.eff_phi4}), an added flat potential region results in an additional second branch of solutions that is classically stable. The resulted effective theory reproduces the main peculiarities of the $(3+1)$-dimensional FLS model, see Fig.\ref{Fig.3D}.   

\begin{center}
  \includegraphics[width=0.7\textwidth]{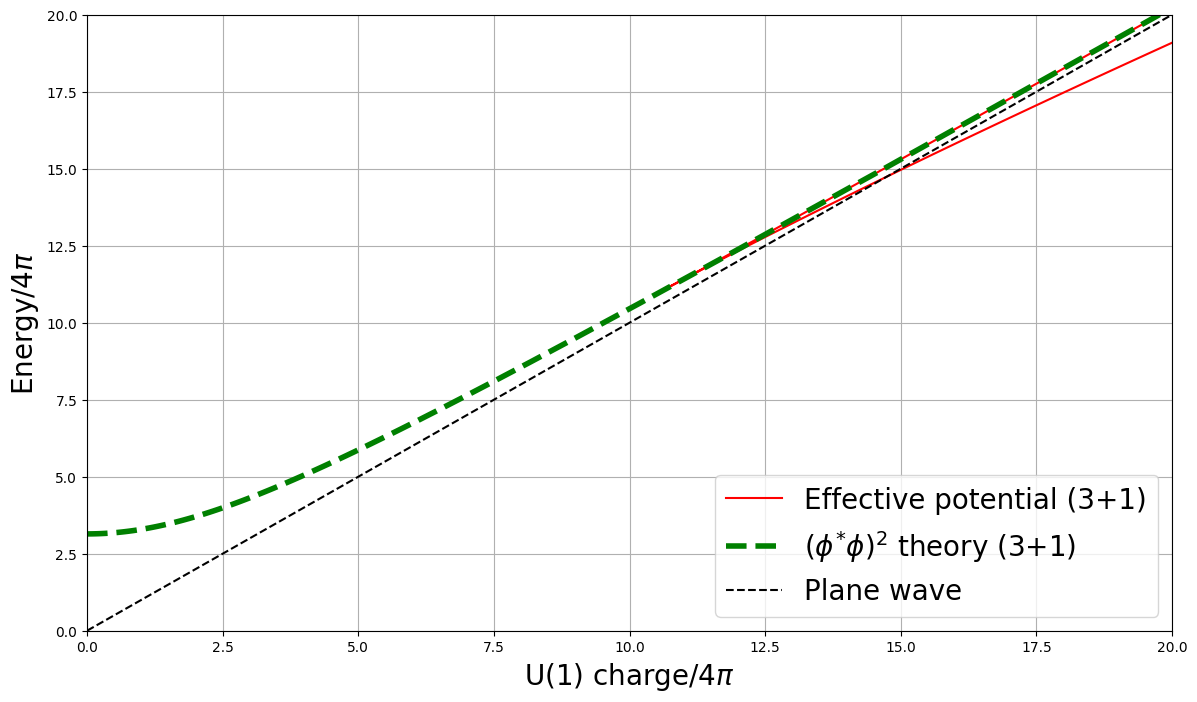}
\captionof{figure}{The energies of non-topological solitons for the $(\phi^{\ast}\phi)^{2}$ theory and model with the effective potential vs. their $U(1)$ charge are plotted for the $(3+1)$-dimensional theory.}\label{Fig.eff_phi4}
\end{center}

\subsection{Parabolic piece-wise potential}
When applied to $(3+1)$-dimension space-time model calculations similar to those performed in Sec.\ref{parabolic1+1} resulted in

\begin{equation}\label{eq3.12}
    \frac{Q}{4\pi} = R^{2}\omega\frac{m^2v^2}{2h^2}\left(\frac{m_{\phi}^{2}(1+R\sqrt{m_{\phi}^{2}-\omega^{2}})}{\omega^{2}\sqrt{m_{\phi}^{2}-\omega^{2}}} \right)
\end{equation}

\begin{equation} \label{eq3.13}
    \frac{E}{4\pi} = \frac{\omega Q}{4 \pi}+ \frac{R^{3}}{3}m_{\phi}^{2}\frac{m^2v^2}{2h^2}
\end{equation}
where $R=\frac{1}{\omega}\left(\pi-\arctan{\left(-\frac{\omega}{\sqrt{m_{\phi}^{2}-\omega^{2}}}\right)} \right)$.

Now the relation (\ref{Legendre}) can be checked for Eqs.(\ref{eq3.12},\ref{eq3.13}).

For the comparison we plotted soliton energy as the function of it's $U(1)$ charge for parabolic piece-wise and effective potentials in Fig.(\ref{Fig.3D}). The difference can be seen in the values of critical parameters of theories and explicitly if the dependence of both integral quantities on $\omega$ is taken into account.

\subsection{Comparison with numerical results}

In this section, we are interested in studying solitons of the FLS model in $(3+1)$ dimensions in comparison to ones from effective potential (\ref{eq3.15}). As can be seen from Eq.(\ref{eq.mot.3+1}) and Fig.(\ref{Fig.3D}), the main differences from the $(1+1)$-dimensional FLS model are the presence of the dissipative term and having two branches of solutions \cite{shnir_2018, PhysRevD.13.2739}.

\begin{equation}\label{eq.mot.3+1}
\begin{cases}
    & \partial_{r}\left(r^2 \partial_{r}\phi \right) = r^{2}\left[h^2\chi^2\phi - \omega^2\phi \right]\\
    & \partial_{r}\left(r^2 \partial_{r}\chi \right) = 
 2 r^{2}\left[h^2|\phi|^2\chi + m^2(\chi^2-v^2)^2 \right]
\end{cases}
\end{equation}

\begin{center}
  \includegraphics[width=0.7\textwidth]{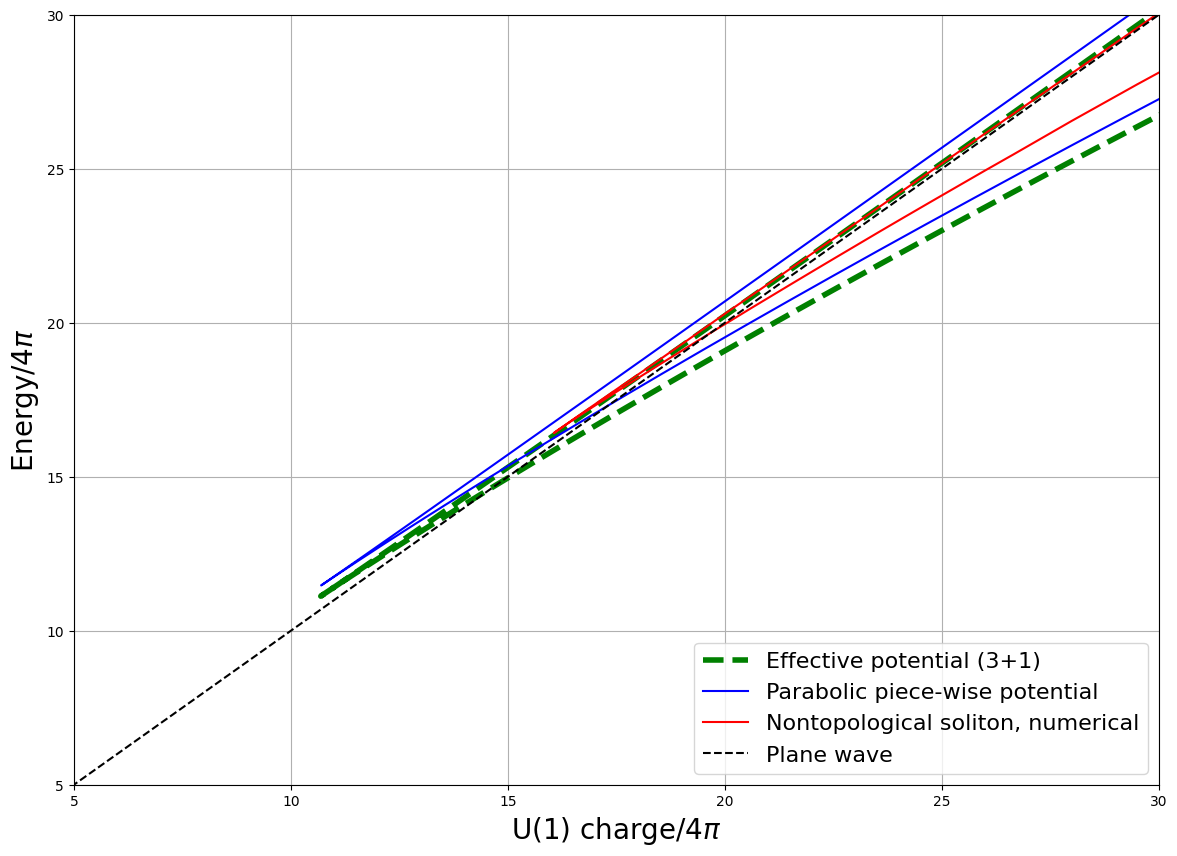}
\captionof{figure}{The $E(Q)$ plots for the FLS model non-topological solitons, effective, and parabolic piece-wise potentials Q-balls are shown for the $(3+1)$-dimensional theory.}\label{Fig.3D}
\end{center}

Fig.(\ref{Fig.3D}) shows that effective potential is better at resurging the results of the original model Eq.(\ref{eq3.1}). At large $U(1)$ charge effective theory gives right asymptotic behavior for both branches of solutions. For the bottom branch the analytically predicted $E \sim Q^{\frac{3}{4}}$ asymptotic behavior is restored in the full and reduced EFT. In accordance with \cite{shnir_2018, PhysRevD.13.2739}, in three spatial dimensions, the gradient term of the Lagrangian has a large contribution to the energy functional, and effective action is required for more precise calculations.

\section{Q-balls within the domain wall}\label{domain walls}

The method of effective potential is inappropriate for the description of topological configurations of the Friedberg-Lee-Sirlin model in which case gradient term is crucial. For large $U(1)$ charge of the field $\phi$ both non-topological and topological solutions behave remarkably similarly, and effective potential is a reliable instrument of reproducing the integral characteristics of the theory. For small charges description of the model by effective potential (\ref{eq3.15}) is not appropriate. However, in this case one can use the perturbation theory in the background of the domain wall.    

The main peculiarity of topological configurations of the FLS model is $\chi \rightarrow -\chi$ symmetry. In order to use perturbation theory for topological configurations, we treated field $\phi$ as a constant background field and compared FLS potential ($\ref{pot}$) to kink potential 

\begin{equation}\label{4.10}
V_{k}(\chi)= \frac{m^{2}}{2}(\chi^{2}-(v^{'})^{2})^{2} \leftrightarrow V(|\phi|^2,\chi)=h^{2}\vert \phi \vert^{2} \chi^{2} + \frac{m^{2}}{2}(\chi^{2}-v^{2})^{2}
\end{equation}


The matching can be explicitly seen from the equation of motion of the field $\chi$

\begin{equation}\label{4.11}
    \chi \left(\frac{h^2}{m^2}|\phi|^2 + (\chi^2-v^2)-(\chi^2-(v^{'})^2)\right)=0
\end{equation}

In terms of Eq.(\ref{4.11}) a deviation of original field $\chi$ from kink configuration is also studied by application of condition (\ref{eq3.5}). 

Through these calculations, one can define $v^{'}=\sqrt{v^{2}-\frac{h^2}{m^2}|\phi|^2}$, and solution of

\begin{equation}\label{4.12}
    \chi^{''}(x)=2h^2|\phi|^2\chi + 2m^2(\chi^2-v^2)\chi=2m^2\left(\chi^2 - \left(v^{'}\right)^{2} \right)\chi
\end{equation}

will take the form of

\begin{equation}\label{eq4.13}
    \chi(|\phi|^{2}) = \sqrt{v^{2}-\frac{h^2}{m^2}|\phi|^2}\tanh{\left( mx\sqrt{v^{2}-\frac{h^2}{m^2}|\phi|^2} \right)}
\end{equation}
and can be used to integrate out the heavy field $\chi$.

Let us revive Sec.\ref{1+1} and reproduce the effective potential using Lagrangian instead of the equations of motion. It can be seen that a series of transformations $\partial_{\mu}\chi\partial^{\mu}\chi \rightarrow -\chi\partial_{\mu}\partial^{\mu}\chi$, $\partial_{\mu}\partial^{\mu}\chi \rightarrow -\frac{\partial V(|\phi|^2,\chi)}{\partial \chi}$ being applied to FLS Lagrangian results in 

\begin{equation}\label{effective lagrangian}
    \mathcal{L}=\partial_{\mu}\phi^{\ast}\partial^{\mu}\phi + \frac{m^2}{2}\left(\chi^4(|\phi|^2) - v^4 \right)
\end{equation}
which leads to the effective potential in the presence of an inhomogeneous background field. A close look at the Eq.(\ref{effective lagrangian}) shows that when applied to Sec.\ref{effective} it provides the same form of $V_{eff}$. Substituting field $\chi$ as Eq.(\ref{eq3.6}) and as $\chi=0$ will lead to potentials $V_{1}$ and $V_{2}$ with only difference that the transition can be done in a continuous way.

\begin{equation}\label{lagrangian transrormed}
    \mathcal{L}=\partial_{\mu}\phi^{\ast}\partial^{\mu}\phi +\frac{m^2}{2}(\chi^{4}(|\phi|^{2})-v^{4})=\partial_{\mu}\phi^{\ast}\partial^{\mu}\phi +\frac{m^2}{2}\left( (v^{2}-\frac{h^2}{m^2}|\phi|^2)^{2}\tanh^{4}{\left( mx\sqrt{v^{2}-\frac{h^2}{m^2}|\phi|^2} \right)} -v^{4}\right)
\end{equation}
when $|\phi|\leq \frac{mv}{h}$, otherwise

\begin{equation}\label{lagrangian transrormed 2}
    \mathcal{L}=\partial_{\mu}\phi^{\ast}\partial^{\mu}\phi - \frac{m^2v^4}{2}
\end{equation}

Eqs.(\ref{lagrangian transrormed},\ref{lagrangian transrormed 2}) can only be solved numerically and result in a qualitatively good approximation of the FLS model for topological configurations.

\begin{center}
  \includegraphics[width=0.7\textwidth]{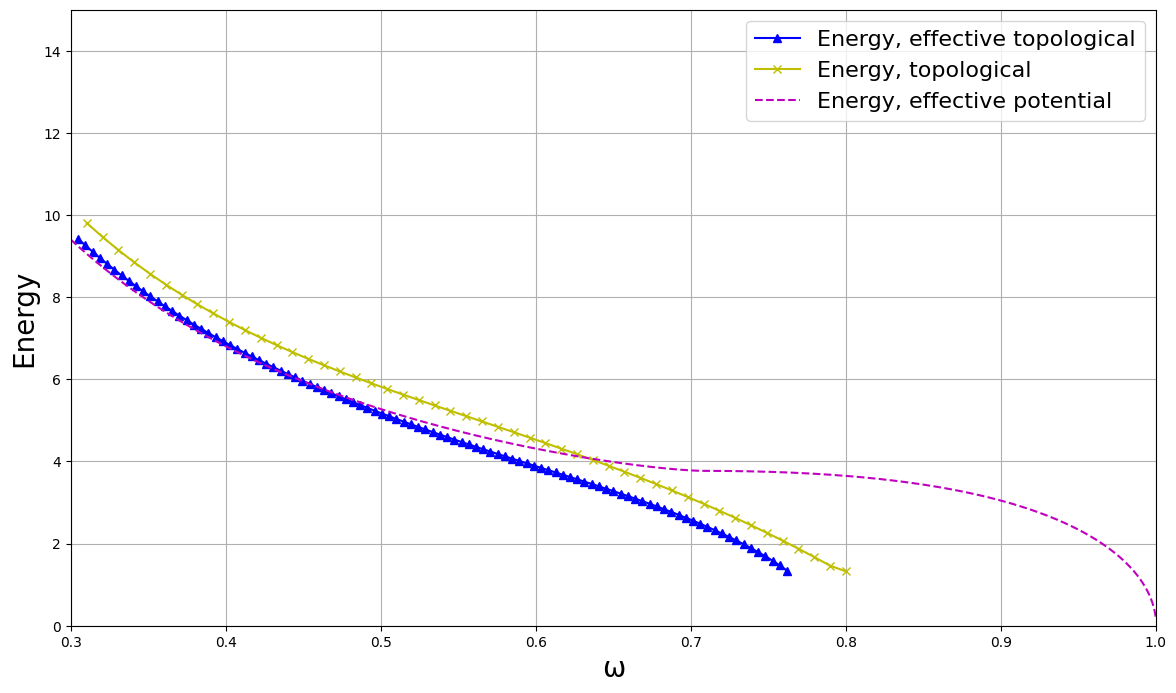}
  \includegraphics[width=0.7\textwidth]{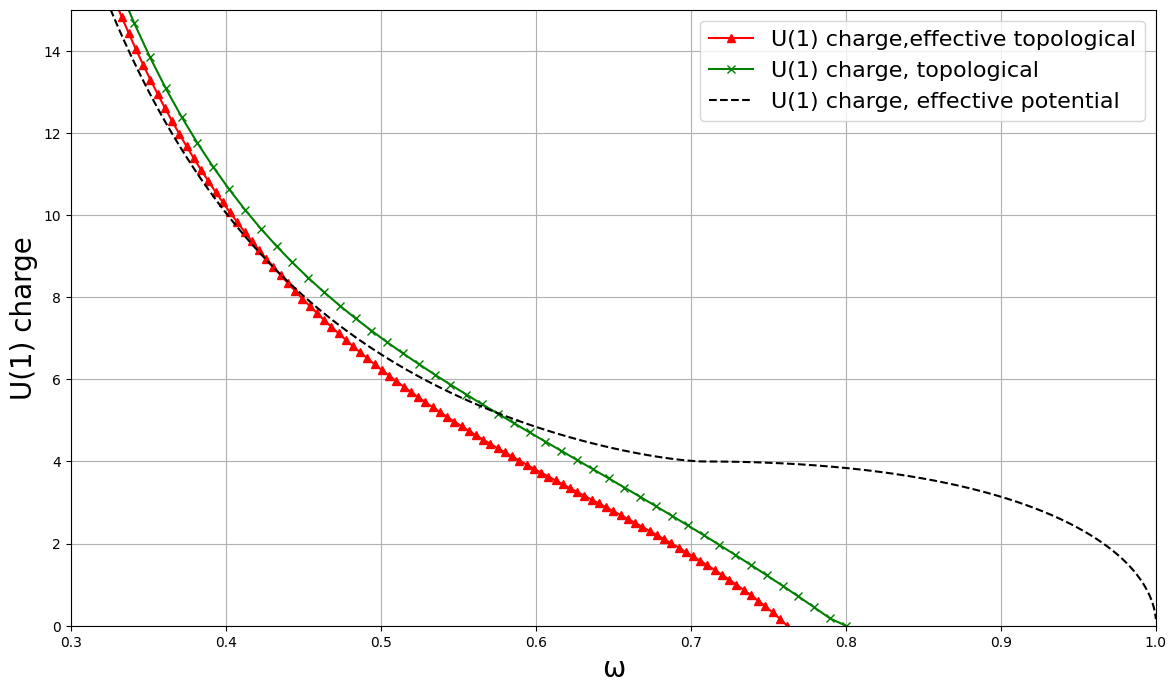}
\captionof{figure}{The energies (top figure) and $U(1)$ charges (bottom figure) of topological configurations of the FLS model, effective potential Q-balls, and Q-balls within the domain wall are plotted as functions of parameter $\omega$. The upper bound in $\omega$ for topological configurations of the FLS model and Q-balls within domain walls appears due to the existence of a bosonic bound state on kink.}\label{Fig.5.1}
\end{center}

\begin{center}
  \includegraphics[width=0.7\textwidth]{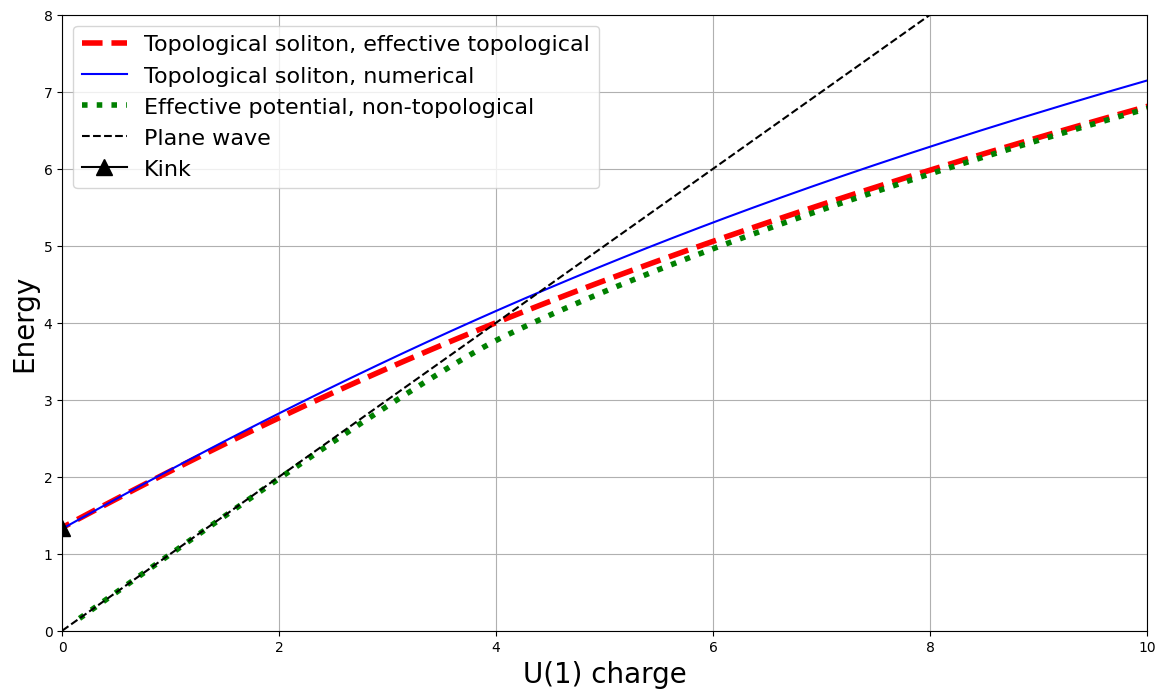}
\captionof{figure}{The energies of the FLS topological configurations, effective potential Q-balls, and Q-balls within domain wall vs. their $U(1)$ charge.}\label{Fig.5.2}
\end{center}

Information about topological configurations of the Friedberg-Lee-Sirlin model could be extracted from the Eq.(\ref{lagrangian transrormed}) by using the perturbation theory. Formal expansion of the Eq.(\ref{eq4.13}) in Taylor series requires condition $\frac{h^2 |\phi|^2}{m^2 v^2} \ll 1$, therefore

\begin{equation}\label{eq4.14}
    \chi \approx v\tanh{(mvx)} + ...
\end{equation}

Equation of motion of theory (\ref{lagrangian transrormed}) in the zeroth-order of perturbation theory can be written by substituting Eq.(\ref{eq4.14}) into the upper equation of Eqs.(\ref{eq3.2}) as follows using ansatz (\ref{eq3.3})

\begin{equation}\label{eq4.16}
    f^{''}(x) + \left[(\omega^{2}-m_{\phi}^{2})+\frac{m_{\phi}^{2}}{\cosh^{2}{(mvx)}}\right]f(x)=0 
\end{equation}
with the lowest frequency solution (see App.\ref{trapping})

\begin{equation}\label{eq4.17}
\omega_{0}^{2}=\frac{m^2v^2}{2}\left( \sqrt{1+\frac{4h^2}{m^2}}-1 \right),\qquad f(x)=\frac{A}{\cosh^{\frac{\sqrt{m_{\phi}^{2}-\omega_{0}^{2}}}{mv}}{(mvx)}}
\end{equation}
which is a bound state of $\phi$-bosons on the domain wall. Even though the calculation were performed in linear approximation the result appears as a non-perturbative effect.

\begin{center}
  \includegraphics[width=0.7\textwidth]{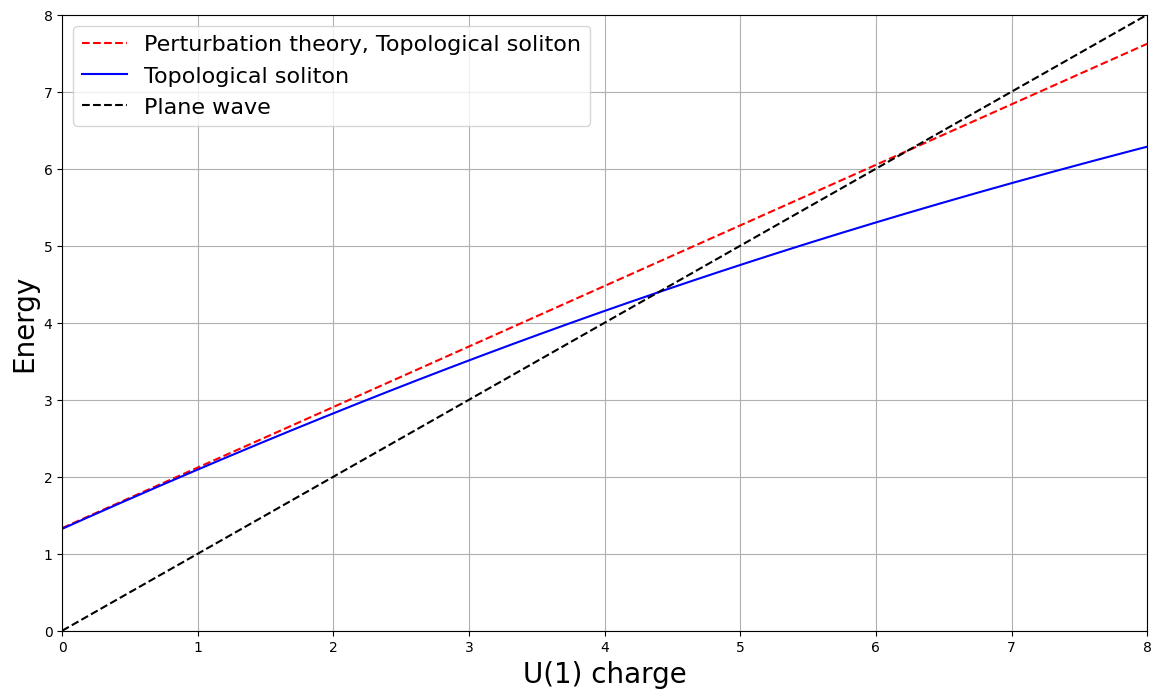}
\captionof{figure}{The comparison of the $E(Q)$ characteristics between topological configurations of the FLS model and bosonic bound state on kink with an arbitrary number of particles is provided.}\label{Fig.6}
\end{center}

Now we can perform integration of solution Eq.(\ref{A1.2}) to obtain energy and $U(1)$ charge

\begin{equation}\label{eq4.18}
    Q = 2\omega_{0}A^{2}\int_{-\infty}^{\infty}dxf^{2}(x) 
\end{equation}

\begin{equation}\label{eq4.19}
    E = \frac{4}{3}mv^{3} + \omega_{0}Q
\end{equation}
for which the following relation $\frac{d E}{d Q}=\omega_{0}$ is fulfilled. The $E(Q)$ characteristics for bosons bound state on kink and topological configurations of the FLS model plotted on Fig.(\ref{Fig.6}) arise new interesting questions. Firstly, the most noticeable change is in transformation of the asymptotic vacuum. As well, topological structure of domain walls reshapes the vacuum making shift in energy by kink's mass. In contrast with non-topological solitons that are compared to plane waves in term of quantum mechanical stability, topological configurations are absolutely stable. One can see that at large charges we cannot restrict ourselves to the zeroth order of the perturbation theory, back-reaction of the bound state on kink must be taken into account (see Sec.\ref{revisited}). Nonetheless, both topological field configurations and configuration (\ref{eq4.17}) might be considered as separate states. Since, by definition, seeking for soliton solutions means studying the true vacuum of the theory at a fixed symmetry charge (Noether or topological), it is not surprising that we have become interested in the mechanism of localization of bosons bounded on kink into a energetically more preferable soliton configuration. This issue may be revised in the quantum theory of solitons. For example, it is interesting to reproduce the results of \cite{DVALI2015338} in the $(1+1)$-dimensional FLS model. 

The results of this section are in agreement with numerical results obtained for topological configurations. Another non-trivial issue is the form of Eq.(\ref{lagrangian transrormed}) as a polynomial function of field $\phi$ which is crucial to explicit understanding of formation of Q-ball. It can be derived in a series expansion of Eq.(\ref{eq4.13}) in spatial coordinate (see App.\ref{phi^6}). 

\subsection{ Q-ball on kink revisited }\label{revisited}

As it can be seen from the analysis above topological configuration of model Eq.(\ref{eq3.1}) with zero $U(1)$ charge is well described as Q-ball on scalar kink. Similar models with fermions and kink was studied in details in \cite{PhysRevD.13.3398, PhysRevD.78.025014, Rubakov:2002fi}. In the Friedberg-Lee-Sirlin model in contrast to referred studies of fermion fields coupled to scalar field by Yukawa interaction, we observed the absence of zeroth mode and severe modification of the vaccum of the theory. According to \cite{PhysRevD.100.105003}, non-zeroth mode bound state of Q-ball localized on kink implies backreaction on the profile of the kink in higher orders of perturbation theory. 

\begin{equation}\label{eq4.20}
    \chi \approx v\tanh (m v x)+\frac{\left(\frac{A}{\cosh^{\frac{\sqrt{m_{\phi}^{2}-\omega_{0}^{2}}}{mv}}{(mvx)}} \right)^2 \left(-h^2 m v x+h^2 m v x \tanh ^2(m v x)-h^2 \tanh (m v x)\right)}{2 m^2 v}+...
\end{equation}

From both equation Eq.(\ref{eq4.20}) and numerical analysis it can be seen that back-reaction of the localization of Q-ball on static kink does not results in kink-antikink oscillations as it was in \cite{PhysRevD.100.105003}. It is worth to mention that bound state Eq.(\ref{eq4.17}) is not affected by Pauli's excluison principle, and could be filled with numerous boson particles. 

\begin{center}
  \includegraphics[width=0.7\textwidth]{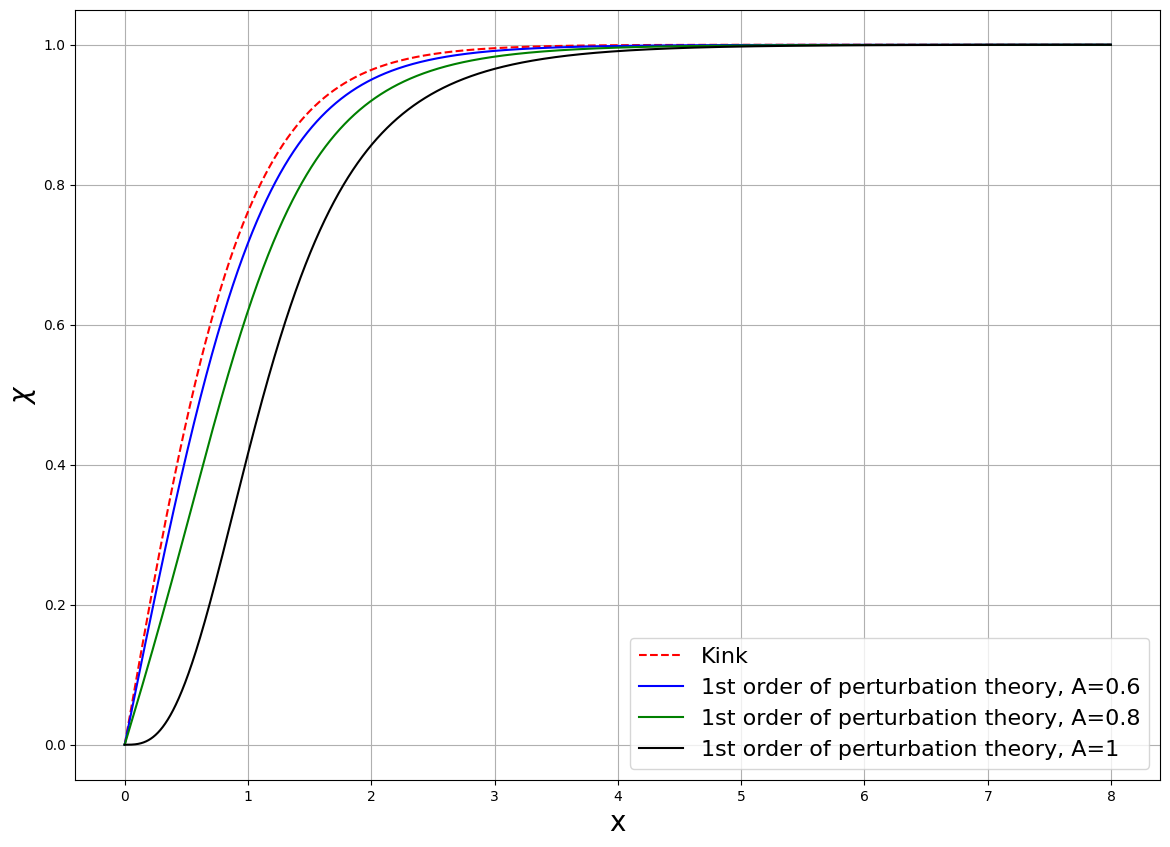}
\captionof{figure}{The profiles of kink and field $\chi$ in the first order of perturbation theory Eq.(\ref{eq4.20}) for different values of amplitude.}\label{Fig.7}
\end{center}

\section{Outlook}\label{outlook}

In this paper, we applied EFT methods to analyze classical field theory solutions. We looked specifically 
at how integrating out the real field in the FLS model affects on the potential of the complex field. The effective potential was constructed by assuming the FLS model's parameter hierarchy. The existence of Q-balls was allowed by the resulting simplified one scalar field potential. Compared to the non-topological solitons of the original theory, Q-balls from the EFT showed reproduce the integral characteristics of the theory in both $(1+1)$ and $(3+1)$ dimensions. The presence of the topological field configurations in the FLS model opened up a question about the possibility of constructing an EFT for the case of an inhomogeneous background. The new EFT is a theory with broken Lorentz symmetry while still allowing for Q-balls that resembled topological configurations of the FLS theory. In addition, by using the perturbation theory in this new EFT, we clarified interpretation of the bound state of bosons on the domain wall. This result was reaffirmed by numerical calculations.      


Solitons in the classical field theory have been found to be useful in various phenomenological models in cosmology, particle physics, etc. Typically, Q-balls appear in supersymmetric theories \cite{KUSENKO199846, ENQVIST1998309, osti_21325400} contrary to solitons of the FLS model. An effective theory developed in this work could be useful for implementing previously developed Q-ball formalism to the EFT Q-balls in the FLS model for phase transitions and dark matter in cosmology \cite{PhysRevD.87.083528, SPECTOR1987103, Troitsky_2016, KUSENKO199846, KUSENKO199726}, particle physics \cite{DVALI199899}, etc. The EFT potentials with flat directions are themselves of interest in early Universe inflation theories \cite{LEE1986181}. Recent research provides observations of nHz gravitational waves (GW) \cite{Agazie_2023, antoniadis2023second, Reardon_2023, Xu_2023} revives the search for sources capable of producing these extreme GW. A discussion of the role of Q-balls in the formation of GW can be found in \cite{PhysRevLett.127.181601, Kasuya_2023, kawasaki2023enhancement}. An oscillon is another type of localized lumps of the classical field \cite{osti_4051808, Bogolyubsky:1976nx, PhysRevD.49.2978}. The main difference between non-topological solitons (or Q-balls) and oscillons is the absence of unbroken internal symmetry. However, approximate conservation of the charge stabilize the solution, resulting in oscillons being dissipative yet long-living objects \cite{Fodor:2019ftc, Zhang_2020}. If both scalar fields are made real in the FLS model, oscillons can form. Results of the Sec.\ref{effective} can provide theory with effectively flat potential that is suitable for the study of oscillons. It may be of research interest due to the proposed role of oscillons in cosmology (see references in \cite{EFT_large_oscillons}).              

The topological structure of the vacuum is another non-trivial aspect of constructing an EFT in the context of studying Q-balls. As previously discussed, effective theory (\ref{eq3.14}) was shown to reproduce the integral characteristics of the FLS model solitons at large charges. Taking into account the topological configurations of the original theory, this effective potential was unsuitable to study Q-balls within a domain wall. By constructing an effective theory in the presence of an inhomogeneous real field, this issue was resolved. Our method, combined with the perturbation theory, made it possible to analyze the condensation of bosons on the domain wall as well as the rearrangement of the theory's vacuum. As a conclusion, we would like to provide a brief discussion of future research on the current issue. Firstly, consistent development requires not only the construction of effective potential but also a method of calculation of an effective action for the FLS theory. Gradient terms of the Lagrangian will be acknowledged as a result of this advancement. In order to have more accurate matching between the EFT Q-balls and FLS solitons, the given step should be resolved. Moreover, constructing an effective action for the original theory not only improves an analysis of the classical solution of field theory but also allows quantum or thermal corrections to be considered. Secondly, the FLS model is a theory of two interacting scalar fields, which makes it possible to take quantum corrections into account and construct Coleman-Weinberg-type effective potential \cite{PhysRevD.7.1888} in this case. Models with soliton solutions can be modified by adding an Abelian gauge field coupled to the original field (or fields) \cite{Loginov:2019sqf, Loginov:2019rwz, Loiko:2019gwk}. The development of EFT for these models may be the subject of future research.

\section{Acknowledgments}\label{acknowledgements}
The authors are grateful to Dmitry Levkov, Anuaruly Oraz, Andrey Shkerin, Yakov Shnir, Mikhail Smolyakov, and Sergey Troitsky for useful discussions and helpful comments on the paper. 
This work was supported by the grant RSF 22-12-00215.

\appendix

\renewcommand{\appendixname}{Appendix}

\section{Bosons on kink}\label{trapping}

In this appendix we will show that like fermions \cite{PhysRevD.100.105003} bosons are also could be localized on kink in $\chi^4$ theory in $(1+1)$-dimensional space-time. We are starting with a corresponding differential equation that describes dynamics of field $\phi$ with interaction with kink within ansatz (\ref{eq3.3})

\begin{equation}\label{A1.1}
    f^{''}(x) + \left[(\omega^{2}-m_{\phi}^{2})+\frac{m_{\phi}^{2}}{\cosh^{2}{(mvx)}}\right]f(x)=0 
\end{equation}

This is also known as bound states problem in modified Pöschl–Teller potential \cite{landau2013quantum}, we obtain solutions of this equation by firstly denoting 

\begin{equation*}
    \xi = \tanh{(mvx)}, \epsilon = \frac{\sqrt{m_{\phi}^{2}-\omega^{2}}}{mv}, s = \frac{-1+\sqrt{1+\frac{4h^2}{m^2}}}{2}
\end{equation*}

After that solution of Eq.(\ref{A1.1}) could be expressed through hypergeometric function as

\begin{equation}\label{A1.2}
    f(x)=A(1-\xi^2)^{\frac{\epsilon}{2}}F\left(\epsilon-s,\epsilon+s+1,\epsilon+1,\frac{1-\xi}{2}\right)
\end{equation}
which due to the arguments given below, is reduced to the Eq.(\ref{eq4.17}).

In order for $f(x)$ to be finite and $f(\infty)=0$ we should keep $\epsilon-s=-n$, where n is the principal quantum number of the corresponding bound state. Only the $0^{\text{th}}$ bound state is of interest since bound states with higher number $n$ experience wave function sign changes. Therefore, we obtain 

\begin{equation}\label{A1.3}
    \omega_{n}^{2}=\left(n+\frac{1}{2}\right)m^2v^2\sqrt{1+\frac{4h^2}{m^2}}-m^2v^2(n^2+n)-\frac{m^2v^2}{2}
\end{equation}

The classical limit on the free parameter of the theory $\omega$ is $\omega \leq m_{\phi}$ results in having only one bound state of bosonic $\phi$ particles on static kink due to Eq.(\ref{A1.3}).

\section{Series expansion in topological configurations}\label{phi^6}

In this appendix, we will illustrate how integrating out field $\chi$ in form of Eq.(\ref{eq4.13}) affects the dynamics of the field $\phi$ in terms of the new polynomial potential $\Tilde{V}(|\phi|)$. These calculations are required for an extensive understanding of the structure of the effective potential in theory (\ref{lagrangian transrormed},\ref{lagrangian transrormed 2}). A convenient way to construct $\Tilde{V}(|\phi|)$ is to start from equation of motion for field $\phi$ 

\begin{equation}\label{B.1}
\begin{split}
    &\partial_{\mu}\partial^{\mu}\phi=\frac{h^2 |\phi|  \left(h^2 |\phi| ^2-m^2 v^2\right) \left(2 m x \sqrt{v^2-\frac{h^2 |\phi| ^2}{m^2}}+\sinh \left(2 m x \sqrt{v^2-\frac{h^2 |\phi| ^2}{m^2}}\right)\right) \tanh ^3\left(m x \sqrt{v^2-\frac{h^2 |\phi| ^2}{m^2}}\right)}{2m^2}\times\\ 
    &\times \frac{\text{sech}^2\left(m x \sqrt{v^2-\frac{h^2 |\phi| ^2}{m^2}}\right)}{2 m^2}
\end{split}
\end{equation}
since RHS of the equation in simply $-\frac{d\Tilde{V}(|\phi|)}{d\phi}$. A better understanding of the physics underlying in Eq.(\ref{B.1}) is given after series expansion in coordinate $x$ is done. Here we restrict ourselves only up to the fourth-order of expansion, and after integration we get



\begin{equation}\label{poly_8}
    \Tilde{V}(|\phi|)= 2h^2 m^4 v^6 x^4 |\phi| ^2 -3 h^4 m^2 v^4 x^4 |\phi| ^4 + 2h^6 v^2 x^4 |\phi| ^6 -\frac{h^8 x^4 |\phi| ^8}{2 m^2}
\end{equation}

\begin{center}
  \includegraphics[width=0.7\textwidth]{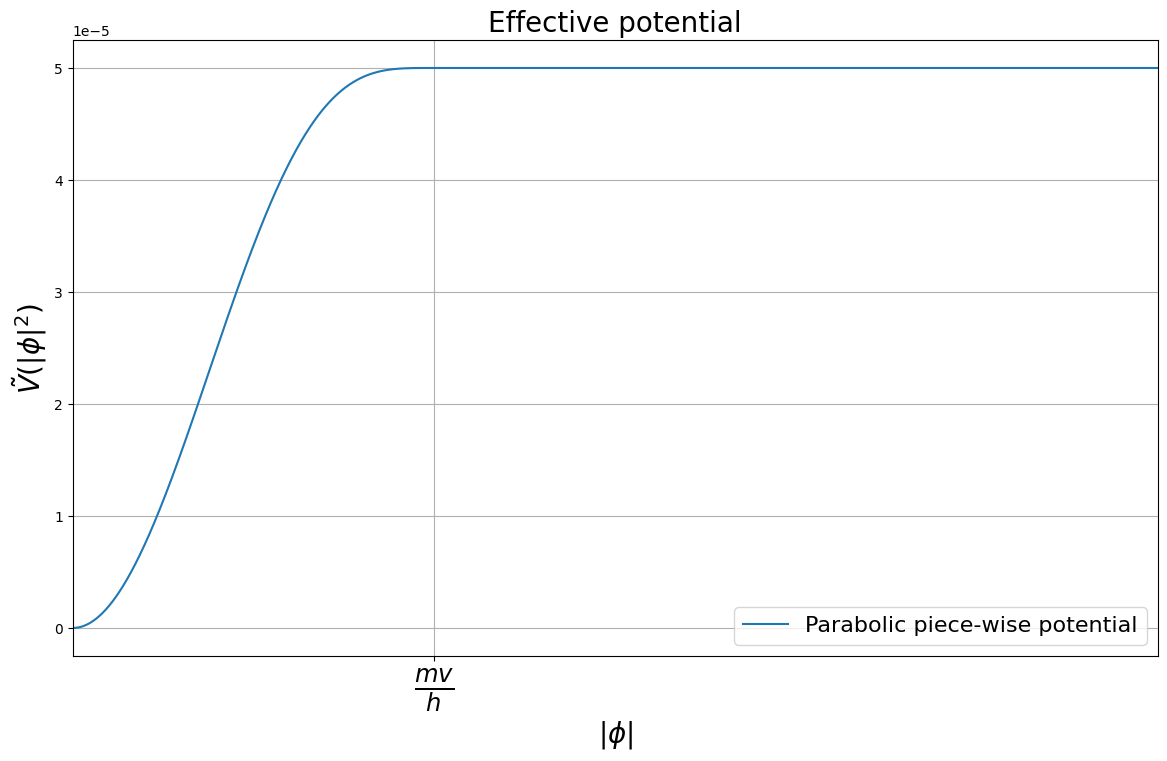}
\captionof{figure}{The effective potential $\Tilde{V}$ gained from the equation of motion (\ref{B.1}) in a series expansion in coordinate variable up to fourth-order and amended with a flat potential as in Sec.\ref{effective}. The profile of the resulted potential (for a given value of coordinate $x=0.1$) allows an analytical prediction of the existence of Q-balls in a theory (\ref{lagrangian transrormed},\ref{lagrangian transrormed 2}).}\label{Fig.8}
\end{center}

After calculations above, we can qualitatively see how $\Tilde{V}(|\phi|)$ is structured. The existence of a Q-ball is provided by the form of the potential, similar to the effective potential (\ref{eq3.15}).

\section{Numerical procedure}\label{numerical}

In this section, we will briefly introduce our numerical method of solving non-linear equations. The first unavoidable step is to transform Lagrangian into dimensionless Lagrangian. When applied to the Friedberg-Lee-Sirlin model, energy $E$ and $U(1)$ charge were transformed to $\Tilde{E}\frac{v}{m}$ and $\frac{\Tilde{Q}}{m^{2}}$ for the $(3+1)$ dimensions and $\Tilde{E}mv$ and $\Tilde{Q}$ for the $(1+1)$ dimensions, where tilde is for dimensionless parameters.

The remaining equations of motion are in the form of

\begin{equation}
    \begin{cases}
        & \nabla^2 f = h^2\chi^2f - \omega^2 f\\
        & \nabla^2 \chi = 2h^2f^2\chi + 2\chi(\chi^2-1)
    \end{cases}
\end{equation}

With boundary conditions

\begin{equation}
    \begin{cases}
        & f^{'}(\infty)=0\\
        & f(\infty)=0\\
        &\chi^{'}(\infty)=0\\
        & \chi(\infty)=1.  
    \end{cases}
\end{equation}

The calculations were performed at fixed parameter $h=1$. As in \cite{PhysRevD.13.2739}, integration is by Runge-Kutta of fourth-order (lattice spacing $\epsilon = 10^{-3}$) with the shooting method of initial conditions. Limitations of shooting parameters could be derived from the analysis of the energy functional of the theory.

Since the non-topological configurations are even, this implies that $\chi^{'}(0)=0$ and $f^{'}(0)=0$). Overall, the following restrictions on initial values of fields are of the form

\begin{equation}
    \begin{cases}
        & \chi(0) \leq \omega\\
        & f(0) \geq \frac{1-\chi^{2}(0)}{\sqrt{2(\omega^{2}-\chi^{2}(0))}}
    \end{cases}
\end{equation}

and for topological configurations (non-zero topological charge implies $\chi(0)=0$ and $\mathcal{Z}_{2}$ symmetry causes $f^{'}(0)=0$)

\begin{equation}
    \begin{cases}
        & \chi^{'}(0)>0 \\
        & f(0) \geq \sqrt{\frac{1-(\chi^{'}(0))^2}{2\omega^2}}
    \end{cases}
\end{equation}

The same method was applied to the theories in which the field $\chi$ was integrated out. In these cases, shooting is much easier due to having only one parameter to shoot-$f(0)$.

\bibliography{biblio}

\begin{thebibliography}{53}%
\makeatletter
\providecommand \@ifxundefined [1]{%
 \@ifx{#1\undefined}
}%
\providecommand \@ifnum [1]{%
 \ifnum #1\expandafter \@firstoftwo
 \else \expandafter \@secondoftwo
 \fi
}%
\providecommand \@ifx [1]{%
 \ifx #1\expandafter \@firstoftwo
 \else \expandafter \@secondoftwo
 \fi
}%
\providecommand \natexlab [1]{#1}%
\providecommand \enquote  [1]{``#1''}%
\providecommand \bibnamefont  [1]{#1}%
\providecommand \bibfnamefont [1]{#1}%
\providecommand \citenamefont [1]{#1}%
\providecommand \href@noop [0]{\@secondoftwo}%
\providecommand \href [0]{\begingroup \@sanitize@url \@href}%
\providecommand \@href[1]{\@@startlink{#1}\@@href}%
\providecommand \@@href[1]{\endgroup#1\@@endlink}%
\providecommand \@sanitize@url [0]{\catcode `\\12\catcode `\$12\catcode
  `\&12\catcode `\#12\catcode `\^12\catcode `\_12\catcode `\%12\relax}%
\providecommand \@@startlink[1]{}%
\providecommand \@@endlink[0]{}%
\providecommand \url  [0]{\begingroup\@sanitize@url \@url }%
\providecommand \@url [1]{\endgroup\@href {#1}{\urlprefix }}%
\providecommand \urlprefix  [0]{URL }%
\providecommand \Eprint [0]{\href }%
\providecommand \doibase [0]{https://doi.org/}%
\providecommand \selectlanguage [0]{\@gobble}%
\providecommand \bibinfo  [0]{\@secondoftwo}%
\providecommand \bibfield  [0]{\@secondoftwo}%
\providecommand \translation [1]{[#1]}%
\providecommand \BibitemOpen [0]{}%
\providecommand \bibitemStop [0]{}%
\providecommand \bibitemNoStop [0]{.\EOS\space}%
\providecommand \EOS [0]{\spacefactor3000\relax}%
\providecommand \BibitemShut  [1]{\csname bibitem#1\endcsname}%
\let\auto@bib@innerbib\@empty
\bibitem [{\citenamefont {Burgess}(2020)}]{Burgess:2020tbq}%
  \BibitemOpen
  \bibfield  {author} {\bibinfo {author} {\bibfnamefont {C.~P.}\ \bibnamefont
  {Burgess}},\ }\href {https://doi.org/10.1017/9781139048040} {\emph {\bibinfo
  {title} {{Introduction to Effective Field Theory}}}}\ (\bibinfo  {publisher}
  {Cambridge University Press},\ \bibinfo {year} {2020})\BibitemShut {NoStop}%
\bibitem [{\citenamefont {Manohar}\ and\ \citenamefont
  {Wise}(2000)}]{manohar_wise_2000}%
  \BibitemOpen
  \bibfield  {author} {\bibinfo {author} {\bibfnamefont {A.~V.}\ \bibnamefont
  {Manohar}}\ and\ \bibinfo {author} {\bibfnamefont {M.~B.}\ \bibnamefont
  {Wise}},\ }\href {https://doi.org/10.1017/CBO9780511529351} {\emph {\bibinfo
  {title} {Heavy Quark Physics}}},\ Cambridge Monographs on Particle Physics,
  Nuclear Physics and Cosmology\ (\bibinfo  {publisher} {Cambridge University
  Press},\ \bibinfo {year} {2000})\BibitemShut {NoStop}%
\bibitem [{\citenamefont {Shifman}\ and\ \citenamefont
  {Voloshin}(1988)}]{Shifman:1987rj}%
  \BibitemOpen
  \bibfield  {author} {\bibinfo {author} {\bibfnamefont {M.~A.}\ \bibnamefont
  {Shifman}}\ and\ \bibinfo {author} {\bibfnamefont {M.~B.}\ \bibnamefont
  {Voloshin}},\ }\bibfield  {title} {\bibinfo {title} {{On Production of d and
  D* Mesons in B Meson Decays}},\ }\href@noop {} {\bibfield  {journal}
  {\bibinfo  {journal} {Sov. J. Nucl. Phys.}\ }\textbf {\bibinfo {volume}
  {47}},\ \bibinfo {pages} {511} (\bibinfo {year} {1988})}\BibitemShut
  {NoStop}%
\bibitem [{\citenamefont {Kajantie}\ \emph {et~al.}(1996)\citenamefont
  {Kajantie}, \citenamefont {Laine}, \citenamefont {Rummukainen},\ and\
  \citenamefont {Shaposhnikov}}]{PhysRevLett.77.2887}%
  \BibitemOpen
  \bibfield  {author} {\bibinfo {author} {\bibfnamefont {K.}~\bibnamefont
  {Kajantie}}, \bibinfo {author} {\bibfnamefont {M.}~\bibnamefont {Laine}},
  \bibinfo {author} {\bibfnamefont {K.}~\bibnamefont {Rummukainen}},\ and\
  \bibinfo {author} {\bibfnamefont {M.}~\bibnamefont {Shaposhnikov}},\
  }\bibfield  {title} {\bibinfo {title} {{Is There a Hot Electroweak Phase
  Transition at ${m}_{H}\ensuremath{\gtrsim}{m}_{W}$?}},\ }\href
  {https://doi.org/10.1103/PhysRevLett.77.2887} {\bibfield  {journal} {\bibinfo
   {journal} {Phys. Rev. Lett.}\ }\textbf {\bibinfo {volume} {77}},\ \bibinfo
  {pages} {2887} (\bibinfo {year} {1996})}\BibitemShut {NoStop}%
\bibitem [{\citenamefont {Karsch}\ \emph {et~al.}(1996)\citenamefont {Karsch},
  \citenamefont {Neuhaus}, \citenamefont {Patkós},\ and\ \citenamefont
  {Rank}}]{KARSCH1996217}%
  \BibitemOpen
  \bibfield  {author} {\bibinfo {author} {\bibfnamefont {F.}~\bibnamefont
  {Karsch}}, \bibinfo {author} {\bibfnamefont {T.}~\bibnamefont {Neuhaus}},
  \bibinfo {author} {\bibfnamefont {A.}~\bibnamefont {Patkós}},\ and\ \bibinfo
  {author} {\bibfnamefont {J.}~\bibnamefont {Rank}},\ }\bibfield  {title}
  {\bibinfo {title} {{Gauge boson masses in the 3D, SU(2) gauge-Higgs model}},\
  }\href {https://doi.org/https://doi.org/10.1016/0550-3213(96)00224-6}
  {\bibfield  {journal} {\bibinfo  {journal} {Nuclear Physics B}\ }\textbf
  {\bibinfo {volume} {474}},\ \bibinfo {pages} {217} (\bibinfo {year}
  {1996})}\BibitemShut {NoStop}%
\bibitem [{\citenamefont {Lee}\ and\ \citenamefont {Pang}(1992)}]{LEE1992251}%
  \BibitemOpen
  \bibfield  {author} {\bibinfo {author} {\bibfnamefont {T.}~\bibnamefont
  {Lee}}\ and\ \bibinfo {author} {\bibfnamefont {Y.}~\bibnamefont {Pang}},\
  }\bibfield  {title} {\bibinfo {title} {{Nontopological solitons}},\ }\href
  {https://doi.org/https://doi.org/10.1016/0370-1573(92)90064-7} {\bibfield
  {journal} {\bibinfo  {journal} {Physics Reports}\ }\textbf {\bibinfo {volume}
  {221}},\ \bibinfo {pages} {251} (\bibinfo {year} {1992})}\BibitemShut
  {NoStop}%
\bibitem [{\citenamefont {Shnir}(2018)}]{shnir_2018}%
  \BibitemOpen
  \bibfield  {author} {\bibinfo {author} {\bibfnamefont {Y.~M.}\ \bibnamefont
  {Shnir}},\ }\href {https://doi.org/10.1017/9781108555623} {\emph {\bibinfo
  {title} {Topological and Non-Topological Solitons in Scalar Field
  Theories}}},\ Cambridge Monographs on Mathematical Physics\ (\bibinfo
  {publisher} {Cambridge University Press},\ \bibinfo {year}
  {2018})\BibitemShut {NoStop}%
\bibitem [{\citenamefont {Nugaev}\ and\ \citenamefont
  {Shkerin}(2020)}]{Nugaev_2020}%
  \BibitemOpen
  \bibfield  {author} {\bibinfo {author} {\bibfnamefont {E.~Y.}\ \bibnamefont
  {Nugaev}}\ and\ \bibinfo {author} {\bibfnamefont {A.~V.}\ \bibnamefont
  {Shkerin}},\ }\bibfield  {title} {\bibinfo {title} {{Review of Nontopological
  Solitons in Theories with U(1)-Symmetry}},\ }\href
  {https://doi.org/10.1134/s1063776120020077} {\bibfield  {journal} {\bibinfo
  {journal} {Journal of Experimental and Theoretical Physics}\ }\textbf
  {\bibinfo {volume} {130}},\ \bibinfo {pages} {301} (\bibinfo {year}
  {2020})}\BibitemShut {NoStop}%
\bibitem [{\citenamefont {Friedberg}\ \emph {et~al.}(1976)\citenamefont
  {Friedberg}, \citenamefont {Lee},\ and\ \citenamefont
  {Sirlin}}]{PhysRevD.13.2739}%
  \BibitemOpen
  \bibfield  {author} {\bibinfo {author} {\bibfnamefont {R.}~\bibnamefont
  {Friedberg}}, \bibinfo {author} {\bibfnamefont {T.~D.}\ \bibnamefont {Lee}},\
  and\ \bibinfo {author} {\bibfnamefont {A.}~\bibnamefont {Sirlin}},\
  }\bibfield  {title} {\bibinfo {title} {{Class of scalar-field soliton
  solutions in three space dimensions}},\ }\href
  {https://doi.org/10.1103/PhysRevD.13.2739} {\bibfield  {journal} {\bibinfo
  {journal} {Phys. Rev. D}\ }\textbf {\bibinfo {volume} {13}},\ \bibinfo
  {pages} {2739} (\bibinfo {year} {1976})}\BibitemShut {NoStop}%
\bibitem [{\citenamefont {Krylov}\ \emph {et~al.}(2013)\citenamefont {Krylov},
  \citenamefont {Levin},\ and\ \citenamefont {Rubakov}}]{PhysRevD.87.083528}%
  \BibitemOpen
  \bibfield  {author} {\bibinfo {author} {\bibfnamefont {E.}~\bibnamefont
  {Krylov}}, \bibinfo {author} {\bibfnamefont {A.}~\bibnamefont {Levin}},\ and\
  \bibinfo {author} {\bibfnamefont {V.}~\bibnamefont {Rubakov}},\ }\bibfield
  {title} {\bibinfo {title} {{Cosmological phase transition, baryon asymmetry,
  and dark matter $Q$-balls}},\ }\href
  {https://doi.org/10.1103/PhysRevD.87.083528} {\bibfield  {journal} {\bibinfo
  {journal} {Phys. Rev. D}\ }\textbf {\bibinfo {volume} {87}},\ \bibinfo
  {pages} {083528} (\bibinfo {year} {2013})}\BibitemShut {NoStop}%
\bibitem [{\citenamefont {Kunz}\ \emph {et~al.}(2022)\citenamefont {Kunz},
  \citenamefont {Loiko},\ and\ \citenamefont {Shnir}}]{PhysRevD.105.085013}%
  \BibitemOpen
  \bibfield  {author} {\bibinfo {author} {\bibfnamefont {J.}~\bibnamefont
  {Kunz}}, \bibinfo {author} {\bibfnamefont {V.}~\bibnamefont {Loiko}},\ and\
  \bibinfo {author} {\bibfnamefont {Y.}~\bibnamefont {Shnir}},\ }\bibfield
  {title} {\bibinfo {title} {{$U(1)$ gauged boson stars in the
  Einstein-Friedberg-Lee-Sirlin model}},\ }\href
  {https://doi.org/10.1103/PhysRevD.105.085013} {\bibfield  {journal} {\bibinfo
   {journal} {Phys. Rev. D}\ }\textbf {\bibinfo {volume} {105}},\ \bibinfo
  {pages} {085013} (\bibinfo {year} {2022})}\BibitemShut {NoStop}%
\bibitem [{\citenamefont {Rajaraman}(1982)}]{rajaraman1982solitons}%
  \BibitemOpen
  \bibfield  {author} {\bibinfo {author} {\bibfnamefont {R.}~\bibnamefont
  {Rajaraman}},\ }\href {https://books.google.kz/books?id=1XucQgAACAAJ} {\emph
  {\bibinfo {title} {Solitons and Instantons: An Introduction to Solitons and
  Instantons in Quantum Field Theory}}},\ North-Holland personal library\
  (\bibinfo  {publisher} {North-Holland Publishing Company},\ \bibinfo {year}
  {1982})\BibitemShut {NoStop}%
\bibitem [{\citenamefont {Manton}\ and\ \citenamefont
  {Sutcliffe}(2004)}]{manton_sutcliffe_2004}%
  \BibitemOpen
  \bibfield  {author} {\bibinfo {author} {\bibfnamefont {N.}~\bibnamefont
  {Manton}}\ and\ \bibinfo {author} {\bibfnamefont {P.}~\bibnamefont
  {Sutcliffe}},\ }\href {https://doi.org/10.1017/CBO9780511617034} {\emph
  {\bibinfo {title} {Topological Solitons}}},\ Cambridge Monographs on
  Mathematical Physics\ (\bibinfo  {publisher} {Cambridge University Press},\
  \bibinfo {year} {2004})\BibitemShut {NoStop}%
\bibitem [{\citenamefont {Rosen}(2003)}]{10.1063/1.1664693}%
  \BibitemOpen
  \bibfield  {author} {\bibinfo {author} {\bibfnamefont {G.}~\bibnamefont
  {Rosen}},\ }\bibfield  {title} {\bibinfo {title} {{Particlelike Solutions to
  Nonlinear Complex Scalar Field Theories with Positive‐Definite Energy
  Densities}},\ }\href {https://doi.org/10.1063/1.1664693} {\bibfield
  {journal} {\bibinfo  {journal} {Journal of Mathematical Physics}\ }\textbf
  {\bibinfo {volume} {9}},\ \bibinfo {pages} {996} (\bibinfo {year}
  {2003})}\BibitemShut {NoStop}%
\bibitem [{\citenamefont {Coleman}(1985)}]{COLEMAN1985263}%
  \BibitemOpen
  \bibfield  {author} {\bibinfo {author} {\bibfnamefont {S.}~\bibnamefont
  {Coleman}},\ }\bibfield  {title} {\bibinfo {title} {{Q-balls}},\ }\href
  {https://doi.org/https://doi.org/10.1016/0550-3213(85)90286-X} {\bibfield
  {journal} {\bibinfo  {journal} {Nuclear Physics B}\ }\textbf {\bibinfo
  {volume} {262}},\ \bibinfo {pages} {263} (\bibinfo {year}
  {1985})}\BibitemShut {NoStop}%
\bibitem [{\citenamefont {Coleman}\ and\ \citenamefont
  {Weinberg}(1973)}]{PhysRevD.7.1888}%
  \BibitemOpen
  \bibfield  {author} {\bibinfo {author} {\bibfnamefont {S.}~\bibnamefont
  {Coleman}}\ and\ \bibinfo {author} {\bibfnamefont {E.}~\bibnamefont
  {Weinberg}},\ }\bibfield  {title} {\bibinfo {title} {{Radiative Corrections
  as the Origin of Spontaneous Symmetry Breaking}},\ }\href
  {https://doi.org/10.1103/PhysRevD.7.1888} {\bibfield  {journal} {\bibinfo
  {journal} {Phys. Rev. D}\ }\textbf {\bibinfo {volume} {7}},\ \bibinfo {pages}
  {1888} (\bibinfo {year} {1973})}\BibitemShut {NoStop}%
\bibitem [{\citenamefont {Lensky}\ \emph {et~al.}(2001)\citenamefont {Lensky},
  \citenamefont {Gani},\ and\ \citenamefont {Kudryavtsev}}]{Lensky_2001}%
  \BibitemOpen
  \bibfield  {author} {\bibinfo {author} {\bibfnamefont {V.~A.}\ \bibnamefont
  {Lensky}}, \bibinfo {author} {\bibfnamefont {V.~A.}\ \bibnamefont {Gani}},\
  and\ \bibinfo {author} {\bibfnamefont {A.~E.}\ \bibnamefont {Kudryavtsev}},\
  }\bibfield  {title} {\bibinfo {title} {{Domain walls carrying a U(1)
  charge}},\ }\href {https://doi.org/10.1134/1.1420436} {\bibfield  {journal}
  {\bibinfo  {journal} {Journal of Experimental and Theoretical Physics}\
  }\textbf {\bibinfo {volume} {93}},\ \bibinfo {pages} {677} (\bibinfo {year}
  {2001})}\BibitemShut {NoStop}%
\bibitem [{\citenamefont {Derrick}(2004)}]{10.1063/1.1704233}%
  \BibitemOpen
  \bibfield  {author} {\bibinfo {author} {\bibfnamefont {G.~H.}\ \bibnamefont
  {Derrick}},\ }\bibfield  {title} {\bibinfo {title} {{Comments on Nonlinear
  Wave Equations as Models for Elementary Particles}},\ }\href
  {https://doi.org/10.1063/1.1704233} {\bibfield  {journal} {\bibinfo
  {journal} {Journal of Mathematical Physics}\ }\textbf {\bibinfo {volume}
  {5}},\ \bibinfo {pages} {1252} (\bibinfo {year} {2004})}\BibitemShut
  {NoStop}%
\bibitem [{\citenamefont {Rajaraman}\ and\ \citenamefont
  {Weinberg}(1975)}]{PhysRevD.11.2950}%
  \BibitemOpen
  \bibfield  {author} {\bibinfo {author} {\bibfnamefont {R.}~\bibnamefont
  {Rajaraman}}\ and\ \bibinfo {author} {\bibfnamefont {E.~J.}\ \bibnamefont
  {Weinberg}},\ }\bibfield  {title} {\bibinfo {title} {{Internal symmetry and
  the semiclassical method in quantum field theory}},\ }\href
  {https://doi.org/10.1103/PhysRevD.11.2950} {\bibfield  {journal} {\bibinfo
  {journal} {Phys. Rev. D}\ }\textbf {\bibinfo {volume} {11}},\ \bibinfo
  {pages} {2950} (\bibinfo {year} {1975})}\BibitemShut {NoStop}%
\bibitem [{\citenamefont {Heeck}\ and\ \citenamefont
  {Sokhashvili}(2023)}]{Heeck:2023idx}%
  \BibitemOpen
  \bibfield  {author} {\bibinfo {author} {\bibfnamefont {J.}~\bibnamefont
  {Heeck}}\ and\ \bibinfo {author} {\bibfnamefont {M.}~\bibnamefont
  {Sokhashvili}},\ }\bibfield  {title} {\bibinfo {title} {{Revisiting the
  Friedberg\textendash{}Lee\textendash{}Sirlin soliton model}},\ }\href
  {https://doi.org/10.1140/epjc/s10052-023-11710-9} {\bibfield  {journal}
  {\bibinfo  {journal} {Eur. Phys. J. C}\ }\textbf {\bibinfo {volume} {83}},\
  \bibinfo {pages} {526} (\bibinfo {year} {2023})},\ \Eprint
  {https://arxiv.org/abs/2303.09566} {arXiv:2303.09566 [hep-ph]} \BibitemShut
  {NoStop}%
\bibitem [{\citenamefont {Montonen}(1976)}]{MONTONEN1976349}%
  \BibitemOpen
  \bibfield  {author} {\bibinfo {author} {\bibfnamefont {C.}~\bibnamefont
  {Montonen}},\ }\bibfield  {title} {\bibinfo {title} {{On solitons with an
  Abelian charge in scalar field theories: (I) Classical theory and
  Bohr-Sommerfeld quantization}},\ }\href
  {https://doi.org/https://doi.org/10.1016/0550-3213(76)90537-X} {\bibfield
  {journal} {\bibinfo  {journal} {Nuclear Physics B}\ }\textbf {\bibinfo
  {volume} {112}},\ \bibinfo {pages} {349} (\bibinfo {year}
  {1976})}\BibitemShut {NoStop}%
\bibitem [{\citenamefont {Gulamov}\ \emph {et~al.}(2013)\citenamefont
  {Gulamov}, \citenamefont {Nugaev},\ and\ \citenamefont
  {Smolyakov}}]{PhysRevD.87.085043}%
  \BibitemOpen
  \bibfield  {author} {\bibinfo {author} {\bibfnamefont {I.~E.}\ \bibnamefont
  {Gulamov}}, \bibinfo {author} {\bibfnamefont {E.~Y.}\ \bibnamefont
  {Nugaev}},\ and\ \bibinfo {author} {\bibfnamefont {M.~N.}\ \bibnamefont
  {Smolyakov}},\ }\bibfield  {title} {\bibinfo {title} {{Analytic $Q$-ball
  solutions and their stability in a piecewise parabolic potential}},\ }\href
  {https://doi.org/10.1103/PhysRevD.87.085043} {\bibfield  {journal} {\bibinfo
  {journal} {Phys. Rev. D}\ }\textbf {\bibinfo {volume} {87}},\ \bibinfo
  {pages} {085043} (\bibinfo {year} {2013})}\BibitemShut {NoStop}%
\bibitem [{\citenamefont {{Anderson}}\ and\ \citenamefont
  {{Derrick}}(1970)}]{1970JMP....11.1336A}%
  \BibitemOpen
  \bibfield  {author} {\bibinfo {author} {\bibfnamefont {D.~L.~T.}\
  \bibnamefont {{Anderson}}}\ and\ \bibinfo {author} {\bibfnamefont {G.~H.}\
  \bibnamefont {{Derrick}}},\ }\bibfield  {title} {\bibinfo {title} {{Stability
  of Time-Dependent Particlelike Solutions in Nonlinear Field Theories. I}},\
  }\href {https://doi.org/10.1063/1.1665265} {\bibfield  {journal} {\bibinfo
  {journal} {Journal of Mathematical Physics}\ }\textbf {\bibinfo {volume}
  {11}},\ \bibinfo {pages} {1336} (\bibinfo {year} {1970})}\BibitemShut
  {NoStop}%
\bibitem [{\citenamefont {Dvali}\ \emph {et~al.}(2015)\citenamefont {Dvali},
  \citenamefont {Gomez}, \citenamefont {Gruending},\ and\ \citenamefont
  {Rug}}]{DVALI2015338}%
  \BibitemOpen
  \bibfield  {author} {\bibinfo {author} {\bibfnamefont {G.}~\bibnamefont
  {Dvali}}, \bibinfo {author} {\bibfnamefont {C.}~\bibnamefont {Gomez}},
  \bibinfo {author} {\bibfnamefont {L.}~\bibnamefont {Gruending}},\ and\
  \bibinfo {author} {\bibfnamefont {T.}~\bibnamefont {Rug}},\ }\bibfield
  {title} {\bibinfo {title} {{Towards a quantum theory of solitons}},\ }\href
  {https://doi.org/https://doi.org/10.1016/j.nuclphysb.2015.10.017} {\bibfield
  {journal} {\bibinfo  {journal} {Nuclear Physics B}\ }\textbf {\bibinfo
  {volume} {901}},\ \bibinfo {pages} {338} (\bibinfo {year}
  {2015})}\BibitemShut {NoStop}%
\bibitem [{\citenamefont {Jackiw}\ and\ \citenamefont
  {Rebbi}(1976)}]{PhysRevD.13.3398}%
  \BibitemOpen
  \bibfield  {author} {\bibinfo {author} {\bibfnamefont {R.}~\bibnamefont
  {Jackiw}}\ and\ \bibinfo {author} {\bibfnamefont {C.}~\bibnamefont {Rebbi}},\
  }\bibfield  {title} {\bibinfo {title} {{Solitons with fermion number
  \textonehalf{}}},\ }\href {https://doi.org/10.1103/PhysRevD.13.3398}
  {\bibfield  {journal} {\bibinfo  {journal} {Phys. Rev. D}\ }\textbf {\bibinfo
  {volume} {13}},\ \bibinfo {pages} {3398} (\bibinfo {year}
  {1976})}\BibitemShut {NoStop}%
\bibitem [{\citenamefont {Brihaye}\ and\ \citenamefont
  {Delsate}(2008)}]{PhysRevD.78.025014}%
  \BibitemOpen
  \bibfield  {author} {\bibinfo {author} {\bibfnamefont {Y.}~\bibnamefont
  {Brihaye}}\ and\ \bibinfo {author} {\bibfnamefont {T.}~\bibnamefont
  {Delsate}},\ }\bibfield  {title} {\bibinfo {title} {{Remarks on bell-shaped
  lumps: Stability and fermionic modes}},\ }\href
  {https://doi.org/10.1103/PhysRevD.78.025014} {\bibfield  {journal} {\bibinfo
  {journal} {Phys. Rev. D}\ }\textbf {\bibinfo {volume} {78}},\ \bibinfo
  {pages} {025014} (\bibinfo {year} {2008})}\BibitemShut {NoStop}%
\bibitem [{\citenamefont {Rubakov}(2002)}]{Rubakov:2002fi}%
  \BibitemOpen
  \bibfield  {author} {\bibinfo {author} {\bibfnamefont {V.~A.}\ \bibnamefont
  {Rubakov}},\ }\href@noop {} {\emph {\bibinfo {title} {{Classical theory of
  gauge fields}}}}\ (\bibinfo  {publisher} {Princeton University Press},\
  \bibinfo {address} {Princeton, New Jersey},\ \bibinfo {year}
  {2002})\BibitemShut {NoStop}%
\bibitem [{\citenamefont {Klimashonok}\ \emph {et~al.}(2019)\citenamefont
  {Klimashonok}, \citenamefont {Perapechka},\ and\ \citenamefont
  {Shnir}}]{PhysRevD.100.105003}%
  \BibitemOpen
  \bibfield  {author} {\bibinfo {author} {\bibfnamefont {V.}~\bibnamefont
  {Klimashonok}}, \bibinfo {author} {\bibfnamefont {I.}~\bibnamefont
  {Perapechka}},\ and\ \bibinfo {author} {\bibfnamefont {Y.}~\bibnamefont
  {Shnir}},\ }\bibfield  {title} {\bibinfo {title} {{Fermions on kinks
  revisited}},\ }\href {https://doi.org/10.1103/PhysRevD.100.105003} {\bibfield
   {journal} {\bibinfo  {journal} {Phys. Rev. D}\ }\textbf {\bibinfo {volume}
  {100}},\ \bibinfo {pages} {105003} (\bibinfo {year} {2019})}\BibitemShut
  {NoStop}%
\bibitem [{\citenamefont {Kusenko}\ and\ \citenamefont
  {Shaposhnikov}(1998)}]{KUSENKO199846}%
  \BibitemOpen
  \bibfield  {author} {\bibinfo {author} {\bibfnamefont {A.}~\bibnamefont
  {Kusenko}}\ and\ \bibinfo {author} {\bibfnamefont {M.}~\bibnamefont
  {Shaposhnikov}},\ }\bibfield  {title} {\bibinfo {title} {{Supersymmetric
  Q-balls as dark matter}},\ }\href
  {https://doi.org/https://doi.org/10.1016/S0370-2693(97)01375-0} {\bibfield
  {journal} {\bibinfo  {journal} {Physics Letters B}\ }\textbf {\bibinfo
  {volume} {418}},\ \bibinfo {pages} {46} (\bibinfo {year} {1998})}\BibitemShut
  {NoStop}%
\bibitem [{\citenamefont {Enqvist}\ and\ \citenamefont
  {McDonald}(1998)}]{ENQVIST1998309}%
  \BibitemOpen
  \bibfield  {author} {\bibinfo {author} {\bibfnamefont {K.}~\bibnamefont
  {Enqvist}}\ and\ \bibinfo {author} {\bibfnamefont {J.}~\bibnamefont
  {McDonald}},\ }\bibfield  {title} {\bibinfo {title} {{Q-balls and
  baryogenesis in the MSSM}},\ }\href
  {https://doi.org/https://doi.org/10.1016/S0370-2693(98)00271-8} {\bibfield
  {journal} {\bibinfo  {journal} {Physics Letters B}\ }\textbf {\bibinfo
  {volume} {425}},\ \bibinfo {pages} {309} (\bibinfo {year}
  {1998})}\BibitemShut {NoStop}%
\bibitem [{\citenamefont {Tsumagari}(2009)}]{osti_21325400}%
  \BibitemOpen
  \bibfield  {author} {\bibinfo {author} {\bibfnamefont {M.~I.}\ \bibnamefont
  {Tsumagari}},\ }\bibfield  {title} {\bibinfo {title} {{Affleck-Dine dynamics,
  Q-ball formation, and thermalization}},\ }\href
  {https://www.osti.gov/biblio/21325400} {\bibfield  {journal} {\bibinfo
  {journal} {Physical Review. D, Particles Fields}\ }\textbf {\bibinfo {volume}
  {80}} (\bibinfo {year} {2009})}\BibitemShut {NoStop}%
\bibitem [{\citenamefont {Spector}(1987)}]{SPECTOR1987103}%
  \BibitemOpen
  \bibfield  {author} {\bibinfo {author} {\bibfnamefont {D.}~\bibnamefont
  {Spector}},\ }\bibfield  {title} {\bibinfo {title} {{First order phase
  transitions in a sector of fixed charge}},\ }\href
  {https://doi.org/https://doi.org/10.1016/0370-2693(87)90777-5} {\bibfield
  {journal} {\bibinfo  {journal} {Physics Letters B}\ }\textbf {\bibinfo
  {volume} {194}},\ \bibinfo {pages} {103} (\bibinfo {year}
  {1987})}\BibitemShut {NoStop}%
\bibitem [{\citenamefont {Troitsky}(2016)}]{Troitsky_2016}%
  \BibitemOpen
  \bibfield  {author} {\bibinfo {author} {\bibfnamefont {S.}~\bibnamefont
  {Troitsky}},\ }\bibfield  {title} {\bibinfo {title} {{Supermassive
  dark-matter Q-balls in galactic centers?}},\ }\href
  {https://doi.org/10.1088/1475-7516/2016/11/027} {\bibfield  {journal}
  {\bibinfo  {journal} {Journal of Cosmology and Astroparticle Physics}\
  }\textbf {\bibinfo {volume} {2016}}\bibinfo  {number} { (11)},\ \bibinfo
  {pages} {027}}\BibitemShut {NoStop}%
\bibitem [{\citenamefont {Kusenko}(1997)}]{KUSENKO199726}%
  \BibitemOpen
\bibfield  {number} {  }\bibfield  {author} {\bibinfo {author} {\bibfnamefont
  {A.}~\bibnamefont {Kusenko}},\ }\bibfield  {title} {\bibinfo {title} {{Phase
  transitions precipitated by solitosynthesis}},\ }\href
  {https://doi.org/https://doi.org/10.1016/S0370-2693(97)00700-4} {\bibfield
  {journal} {\bibinfo  {journal} {Physics Letters B}\ }\textbf {\bibinfo
  {volume} {406}},\ \bibinfo {pages} {26} (\bibinfo {year} {1997})}\BibitemShut
  {NoStop}%
\bibitem [{\citenamefont {Dvali}\ \emph {et~al.}(1998)\citenamefont {Dvali},
  \citenamefont {Kusenko},\ and\ \citenamefont {Shaposhnikov}}]{DVALI199899}%
  \BibitemOpen
  \bibfield  {author} {\bibinfo {author} {\bibfnamefont {G.}~\bibnamefont
  {Dvali}}, \bibinfo {author} {\bibfnamefont {A.}~\bibnamefont {Kusenko}},\
  and\ \bibinfo {author} {\bibfnamefont {M.}~\bibnamefont {Shaposhnikov}},\
  }\bibfield  {title} {\bibinfo {title} {{New physics in a nutshell, or Q-ball
  as a power plant}},\ }\href
  {https://doi.org/https://doi.org/10.1016/S0370-2693(97)01378-6} {\bibfield
  {journal} {\bibinfo  {journal} {Physics Letters B}\ }\textbf {\bibinfo
  {volume} {417}},\ \bibinfo {pages} {99} (\bibinfo {year} {1998})}\BibitemShut
  {NoStop}%
\bibitem [{\citenamefont {Lee}\ and\ \citenamefont
  {Weinberg}(1986)}]{LEE1986181}%
  \BibitemOpen
  \bibfield  {author} {\bibinfo {author} {\bibfnamefont {K.}~\bibnamefont
  {Lee}}\ and\ \bibinfo {author} {\bibfnamefont {E.~J.}\ \bibnamefont
  {Weinberg}},\ }\bibfield  {title} {\bibinfo {title} {{Tunneling without
  barriers}},\ }\href
  {https://doi.org/https://doi.org/10.1016/0550-3213(86)90150-1} {\bibfield
  {journal} {\bibinfo  {journal} {Nuclear Physics B}\ }\textbf {\bibinfo
  {volume} {267}},\ \bibinfo {pages} {181} (\bibinfo {year}
  {1986})}\BibitemShut {NoStop}%
\bibitem [{\citenamefont {Agazie}\ \emph {et~al.}(2023)\citenamefont {Agazie}
  \emph {et~al.}}]{Agazie_2023}%
  \BibitemOpen
  \bibfield  {author} {\bibinfo {author} {\bibfnamefont {G.}~\bibnamefont
  {Agazie}} \emph {et~al.},\ }\bibfield  {title} {\bibinfo {title} {{The
  NANOGrav 15 yr Data Set: Evidence for a Gravitational-wave Background}},\
  }\href {https://doi.org/10.3847/2041-8213/acdac6} {\bibfield  {journal}
  {\bibinfo  {journal} {The Astrophysical Journal Letters}\ }\textbf {\bibinfo
  {volume} {951}},\ \bibinfo {pages} {L8} (\bibinfo {year} {2023})}\BibitemShut
  {NoStop}%
\bibitem [{\citenamefont {Antoniadis}\ \emph {et~al.}(2023)\citenamefont
  {Antoniadis} \emph {et~al.}}]{antoniadis2023second}%
  \BibitemOpen
  \bibfield  {author} {\bibinfo {author} {\bibfnamefont {J.}~\bibnamefont
  {Antoniadis}} \emph {et~al.},\ }\href@noop {} {\bibinfo {title} {The second
  data release from the european pulsar timing array iii. search for
  gravitational wave signals}} (\bibinfo {year} {2023}),\ \Eprint
  {https://arxiv.org/abs/2306.16214} {arXiv:2306.16214 [astro-ph.HE]}
  \BibitemShut {NoStop}%
\bibitem [{\citenamefont {Reardon}\ \emph {et~al.}(2023)\citenamefont {Reardon}
  \emph {et~al.}}]{Reardon_2023}%
  \BibitemOpen
  \bibfield  {author} {\bibinfo {author} {\bibfnamefont {D.~J.}\ \bibnamefont
  {Reardon}} \emph {et~al.},\ }\bibfield  {title} {\bibinfo {title} {{Search
  for an Isotropic Gravitational-wave Background with the Parkes Pulsar Timing
  Array}},\ }\href {https://doi.org/10.3847/2041-8213/acdd02} {\bibfield
  {journal} {\bibinfo  {journal} {The Astrophysical Journal Letters}\ }\textbf
  {\bibinfo {volume} {951}},\ \bibinfo {pages} {L6} (\bibinfo {year}
  {2023})}\BibitemShut {NoStop}%
\bibitem [{\citenamefont {Xu}\ \emph {et~al.}(2023)\citenamefont {Xu} \emph
  {et~al.}}]{Xu_2023}%
  \BibitemOpen
  \bibfield  {author} {\bibinfo {author} {\bibfnamefont {H.}~\bibnamefont {Xu}}
  \emph {et~al.},\ }\bibfield  {title} {\bibinfo {title} {{Searching for the
  Nano-Hertz Stochastic Gravitational Wave Background with the Chinese Pulsar
  Timing Array Data Release I}},\ }\href
  {https://doi.org/10.1088/1674-4527/acdfa5} {\bibfield  {journal} {\bibinfo
  {journal} {Research in Astronomy and Astrophysics}\ }\textbf {\bibinfo
  {volume} {23}},\ \bibinfo {pages} {075024} (\bibinfo {year}
  {2023})}\BibitemShut {NoStop}%
\bibitem [{\citenamefont {White}\ \emph {et~al.}(2021)\citenamefont {White},
  \citenamefont {Pearce}, \citenamefont {Vagie},\ and\ \citenamefont
  {Kusenko}}]{PhysRevLett.127.181601}%
  \BibitemOpen
  \bibfield  {author} {\bibinfo {author} {\bibfnamefont {G.}~\bibnamefont
  {White}}, \bibinfo {author} {\bibfnamefont {L.}~\bibnamefont {Pearce}},
  \bibinfo {author} {\bibfnamefont {D.}~\bibnamefont {Vagie}},\ and\ \bibinfo
  {author} {\bibfnamefont {A.}~\bibnamefont {Kusenko}},\ }\bibfield  {title}
  {\bibinfo {title} {{Detectable Gravitational Wave Signals from Affleck-Dine
  Baryogenesis}},\ }\href {https://doi.org/10.1103/PhysRevLett.127.181601}
  {\bibfield  {journal} {\bibinfo  {journal} {Phys. Rev. Lett.}\ }\textbf
  {\bibinfo {volume} {127}},\ \bibinfo {pages} {181601} (\bibinfo {year}
  {2021})}\BibitemShut {NoStop}%
\bibitem [{\citenamefont {Kasuya}\ \emph {et~al.}(2023)\citenamefont {Kasuya},
  \citenamefont {Kawasaki},\ and\ \citenamefont {Murai}}]{Kasuya_2023}%
  \BibitemOpen
  \bibfield  {author} {\bibinfo {author} {\bibfnamefont {S.}~\bibnamefont
  {Kasuya}}, \bibinfo {author} {\bibfnamefont {M.}~\bibnamefont {Kawasaki}},\
  and\ \bibinfo {author} {\bibfnamefont {K.}~\bibnamefont {Murai}},\ }\bibfield
   {title} {\bibinfo {title} {{Enhancement of second-order gravitational waves
  at Q-ball decay}},\ }\href {https://doi.org/10.1088/1475-7516/2023/05/053}
  {\bibfield  {journal} {\bibinfo  {journal} {Journal of Cosmology and
  Astroparticle Physics}\ }\textbf {\bibinfo {volume} {2023}}\bibinfo  {number}
  { (05)},\ \bibinfo {pages} {053}}\BibitemShut {NoStop}%
\bibitem [{\citenamefont {Kawasaki}\ and\ \citenamefont
  {Murai}(2023)}]{kawasaki2023enhancement}%
  \BibitemOpen
\bibfield  {number} {  }\bibfield  {author} {\bibinfo {author} {\bibfnamefont
  {M.}~\bibnamefont {Kawasaki}}\ and\ \bibinfo {author} {\bibfnamefont
  {K.}~\bibnamefont {Murai}},\ }\href@noop {} {\bibinfo {title} {Enhancement of
  gravitational waves at q-ball decay including non-linear density
  perturbations}} (\bibinfo {year} {2023}),\ \Eprint
  {https://arxiv.org/abs/2308.13134} {arXiv:2308.13134 [astro-ph.CO]}
  \BibitemShut {NoStop}%
\bibitem [{\citenamefont {Kudryavtsev}(1975)}]{osti_4051808}%
  \BibitemOpen
  \bibfield  {author} {\bibinfo {author} {\bibfnamefont {A.~E.}\ \bibnamefont
  {Kudryavtsev}},\ }\bibfield  {title} {\bibinfo {title} {{Solitonlike
  solutions for a Higgs scalar field}},\ }\bibfield  {journal} {\bibinfo
  {journal} {JETP Lett. (USSR) (Engl. Transl.), v. 22, no. 3, pp. 82-83}\
  }\textbf {\bibinfo {volume} {22}},\ \href
  {https://www.osti.gov/biblio/4051808} {} (\bibinfo {year} {1975})\BibitemShut
  {NoStop}%
\bibitem [{\citenamefont {Bogolyubsky}\ and\ \citenamefont
  {Makhankov}(1976)}]{Bogolyubsky:1976nx}%
  \BibitemOpen
  \bibfield  {author} {\bibinfo {author} {\bibfnamefont {I.~L.}\ \bibnamefont
  {Bogolyubsky}}\ and\ \bibinfo {author} {\bibfnamefont {V.~G.}\ \bibnamefont
  {Makhankov}},\ }\bibfield  {title} {\bibinfo {title} {{On the Pulsed Soliton
  Lifetime in Two Classical Relativistic Theory Models}},\ }\href@noop {}
  {\bibfield  {journal} {\bibinfo  {journal} {JETP Lett.}\ }\textbf {\bibinfo
  {volume} {24}},\ \bibinfo {pages} {12} (\bibinfo {year} {1976})}\BibitemShut
  {NoStop}%
\bibitem [{\citenamefont {Gleiser}(1994)}]{PhysRevD.49.2978}%
  \BibitemOpen
  \bibfield  {author} {\bibinfo {author} {\bibfnamefont {M.}~\bibnamefont
  {Gleiser}},\ }\bibfield  {title} {\bibinfo {title} {{Pseudostable bubbles}},\
  }\href {https://doi.org/10.1103/PhysRevD.49.2978} {\bibfield  {journal}
  {\bibinfo  {journal} {Phys. Rev. D}\ }\textbf {\bibinfo {volume} {49}},\
  \bibinfo {pages} {2978} (\bibinfo {year} {1994})}\BibitemShut {NoStop}%
\bibitem [{\citenamefont {Fodor}(2019)}]{Fodor:2019ftc}%
  \BibitemOpen
  \bibfield  {author} {\bibinfo {author} {\bibfnamefont {G.}~\bibnamefont
  {Fodor}},\ }\emph {\bibinfo {title} {{A review on radiation of oscillons and
  oscillatons}}},\ \href@noop {} {Ph.D. thesis},\ \bibinfo  {school} {Wigner
  RCP, Budapest} (\bibinfo {year} {2019}),\ \Eprint
  {https://arxiv.org/abs/1911.03340} {arXiv:1911.03340 [hep-th]} \BibitemShut
  {NoStop}%
\bibitem [{\citenamefont {Zhang}\ \emph {et~al.}(2020)\citenamefont {Zhang},
  \citenamefont {Amin}, \citenamefont {Copeland}, \citenamefont {Saffin},\ and\
  \citenamefont {Lozanov}}]{Zhang_2020}%
  \BibitemOpen
  \bibfield  {author} {\bibinfo {author} {\bibfnamefont {H.-Y.}\ \bibnamefont
  {Zhang}}, \bibinfo {author} {\bibfnamefont {M.~A.}\ \bibnamefont {Amin}},
  \bibinfo {author} {\bibfnamefont {E.~J.}\ \bibnamefont {Copeland}}, \bibinfo
  {author} {\bibfnamefont {P.~M.}\ \bibnamefont {Saffin}},\ and\ \bibinfo
  {author} {\bibfnamefont {K.~D.}\ \bibnamefont {Lozanov}},\ }\bibfield
  {title} {\bibinfo {title} {{Classical decay rates of oscillons}},\ }\href
  {https://doi.org/10.1088/1475-7516/2020/07/055} {\bibfield  {journal}
  {\bibinfo  {journal} {Journal of Cosmology and Astroparticle Physics}\
  }\textbf {\bibinfo {volume} {2020}}\bibinfo  {number} { (07)},\ \bibinfo
  {pages} {055}}\BibitemShut {NoStop}%
\bibitem [{\citenamefont {Levkov}\ \emph {et~al.}(2022)\citenamefont {Levkov},
  \citenamefont {Maslov}, \citenamefont {Nugaev},\ and\ \citenamefont
  {Panin}}]{EFT_large_oscillons}%
  \BibitemOpen
\bibfield  {number} {  }\bibfield  {author} {\bibinfo {author} {\bibfnamefont
  {D.}~\bibnamefont {Levkov}}, \bibinfo {author} {\bibfnamefont
  {V.}~\bibnamefont {Maslov}}, \bibinfo {author} {\bibfnamefont
  {E.}~\bibnamefont {Nugaev}},\ and\ \bibinfo {author} {\bibfnamefont
  {A.}~\bibnamefont {Panin}},\ }\bibfield  {title} {\bibinfo {title} {{An
  Effective Field Theory for large oscillons}},\ }\href
  {https://doi.org/10.1007/JHEP12(2022)079} {\bibfield  {journal} {\bibinfo
  {journal} {Journal of High Energy Physics}\ }\textbf {\bibinfo {volume}
  {2022}},\ \bibinfo {pages} {79} (\bibinfo {year} {2022})}\BibitemShut
  {NoStop}%
\bibitem [{\citenamefont {Loginov}\ and\ \citenamefont
  {Gauzshtein}(2019{\natexlab{a}})}]{Loginov:2019sqf}%
  \BibitemOpen
  \bibfield  {author} {\bibinfo {author} {\bibfnamefont {A.~Y.}\ \bibnamefont
  {Loginov}}\ and\ \bibinfo {author} {\bibfnamefont {V.~V.}\ \bibnamefont
  {Gauzshtein}},\ }\bibfield  {title} {\bibinfo {title} {{One-dimensional
  soliton system of gauged Q -ball and anti- Q -ball}},\ }\href
  {https://doi.org/10.1103/PhysRevD.99.065011} {\bibfield  {journal} {\bibinfo
  {journal} {Phys. Rev. D}\ }\textbf {\bibinfo {volume} {99}},\ \bibinfo
  {pages} {065011} (\bibinfo {year} {2019}{\natexlab{a}})},\ \Eprint
  {https://arxiv.org/abs/1901.00272} {arXiv:1901.00272 [hep-th]} \BibitemShut
  {NoStop}%
\bibitem [{\citenamefont {Loginov}\ and\ \citenamefont
  {Gauzshtein}(2019{\natexlab{b}})}]{Loginov:2019rwz}%
  \BibitemOpen
  \bibfield  {author} {\bibinfo {author} {\bibfnamefont {A.~Y.}\ \bibnamefont
  {Loginov}}\ and\ \bibinfo {author} {\bibfnamefont {V.~V.}\ \bibnamefont
  {Gauzshtein}},\ }\bibfield  {title} {\bibinfo {title} {{One-dimensional
  soliton system of gauged kink and Q-ball}},\ }\href
  {https://doi.org/10.1140/epjc/s10052-019-7302-6} {\bibfield  {journal}
  {\bibinfo  {journal} {Eur. Phys. J. C}\ }\textbf {\bibinfo {volume} {79}},\
  \bibinfo {pages} {780} (\bibinfo {year} {2019}{\natexlab{b}})},\ \Eprint
  {https://arxiv.org/abs/1906.02447} {arXiv:1906.02447 [hep-th]} \BibitemShut
  {NoStop}%
\bibitem [{\citenamefont {Loiko}\ and\ \citenamefont
  {Shnir}(2019)}]{Loiko:2019gwk}%
  \BibitemOpen
  \bibfield  {author} {\bibinfo {author} {\bibfnamefont {V.}~\bibnamefont
  {Loiko}}\ and\ \bibinfo {author} {\bibfnamefont {Y.}~\bibnamefont {Shnir}},\
  }\bibfield  {title} {\bibinfo {title} {{Q-balls in the $U(1)$ gauged
  Friedberg-Lee-Sirlin model}},\ }\href
  {https://doi.org/10.1016/j.physletb.2019.134810} {\bibfield  {journal}
  {\bibinfo  {journal} {Phys. Lett. B}\ }\textbf {\bibinfo {volume} {797}},\
  \bibinfo {pages} {134810} (\bibinfo {year} {2019})},\ \Eprint
  {https://arxiv.org/abs/1906.01943} {arXiv:1906.01943 [hep-th]} \BibitemShut
  {NoStop}%
\bibitem [{\citenamefont {Landau}\ and\ \citenamefont
  {Lifshitz}(2013)}]{landau2013quantum}%
  \BibitemOpen
  \bibfield  {author} {\bibinfo {author} {\bibfnamefont {L.~D.}\ \bibnamefont
  {Landau}}\ and\ \bibinfo {author} {\bibfnamefont {E.~M.}\ \bibnamefont
  {Lifshitz}},\ }\href@noop {} {\emph {\bibinfo {title} {Quantum mechanics:
  non-relativistic theory}}},\ Vol.~\bibinfo {volume} {3}\ (\bibinfo
  {publisher} {Elsevier},\ \bibinfo {year} {2013})\BibitemShut {NoStop}%
\end{thebibliography}%

\end{document}